\documentclass[twocolumn,twocolappendix]{aastex631}
\usepackage{bm}
\usepackage{amsmath,amsthm,amsfonts,amssymb,amscd}
\usepackage{threeparttable}
\usepackage[]{color}

\shorttitle{Galaxy-halo alignment in Hydrodynamical Simulation}
\shortauthors{Xu, Jing \& Zhao}

\begin{document}
%\title{First Measurements of Baryon Acoustic Oscillation Dips in Galaxy Ellipticity Correlations}

\title{Toward a Physical Understanding of Galaxy-Halo Alignment}

\correspondingauthor{Y.P. Jing}
\email{ypjing@sjtu.edu.cn}

\author[0000-0002-7697-3306]{Kun Xu}
\affil{Department of Astronomy, School of Physics and Astronomy, Shanghai Jiao Tong University, Shanghai, 200240, People’s Republic of China}
\affil{Institute for Computational Cosmology, Department of Physics, Durham University, South Road, Durham DH1 3LE, UK}

\author[0000-0002-4534-3125]{Y.P. Jing}
\affil{Department of Astronomy, School of Physics and Astronomy, Shanghai Jiao Tong University, Shanghai, 200240, People’s Republic of China}
\affil{Tsung-Dao Lee Institute, and Shanghai Key Laboratory for Particle Physics and Cosmology, Shanghai Jiao Tong University, Shanghai, 200240, People’s Republic of China}

\author{Donghai Zhao}
\affil{Key Laboratory for Research in Galaxies and Cosmology, Shanghai Astronomical Observatory, Shanghai, 200030, People's Republic of
China}
\affil{Department of Astronomy, School of Physics and Astronomy, Shanghai Jiao Tong University, Shanghai, 200240, People’s Republic of China}

\begin{abstract}
We investigate the alignment of galaxy and halo orientations using the TNG300-1 hydrodynamical simulation. Our analysis reveals that the distribution of the 2D misalignment angle $\theta_{\rm{2D}}$ can be well described by a truncated shifted exponential (TSE) distribution with only {\textit{one}} free parameter across different redshifts and galaxy/halo properties. We demonstrate that the galaxy-ellipticity (GI) correlations of galaxies can be reproduced by perturbing halo orientations with the obtained $\theta_{\rm{2D}}$ distribution, with only a small bias ($<3^{\circ}$) possibly arising from unaccounted couplings between $\theta_{\rm{2D}}$ and other factors. We find that both the 2D and 3D misalignment angles $\theta_{\rm{2D}}$ and $\theta_{\rm{3D}}$ decrease with ex situ stellar mass fraction $F_{\rm{acc}}$, halo mass $M_{\rm{vir}}$ and stellar mass $M_{*}$, while increasing with disk-to-total stellar mass fraction $F_{\rm{disk}}$ and redshift. These dependences are in good agreement with our recent observational study based on the BOSS galaxy samples. Our results suggest that $F_{\rm{acc}}$ is a key factor in determining the galaxy-halo alignment. Grouping galaxies by $F_{\rm{acc}}$ nearly eliminates the dependence of $\theta_{\rm{3D}}$ on $M_{\rm{vir}}$ for all three principle axes, and also reduces the redshift dependence. For $\theta_{\rm{2D}}$, we find a more significant redshift dependence than for  $\theta_{\rm{3D}}$ even after controlling $F_{\rm{acc}}$, which may be attributed to the evolution of galaxy and halo shapes. Our findings present a valuable model for observational studies and enhance our understanding of galaxy-halo alignment. 
\end{abstract}
%% Keywords should appear after the \end{abstract} command. 
%% See the online documentation for the full list of available subject
%% keywords and the rules for their use.

\section{Introduction} \label{sec:1}
The intrinsic correlation between galaxy shape and the large-scale structure of the Universe has been discovered for a long time, know as "intrinsic alignment" (IA) \citep{2000ApJ...543L.107P,2002MNRAS.333..501B,2004MNRAS.353..529H,2006MNRAS.367..611M,2006MNRAS.369.1293Y,2009ApJ...694..214O,2009ApJ...694L..83O,2013ApJ...770L..12L,2015MNRAS.450.2195S,2022MNRAS.514.1077R,2023ApJ...943..128S,2023arXiv230204230X}. It has long be known to be a main contamination to weak lensing measurement that can result in biased constraints on the cosmological parameters \citep{2000ApJ...545..561C,2004PhRvD..70f3526H,2007MNRAS.381.1197H,2012MNRAS.424.1647K,2016MNRAS.456..207K}. Recently, several studies suggested that intrinsic alignment (IA) could also serve as a promising cosmological probe in future surveys \citep{2013JCAP...12..029C,2015JCAP...10..032S,2016PhRvD..94l3507C,2018JCAP...08..014K,2020MNRAS.493L.124O,2022PhRvD.106d3523O}, and first attempts have been made in measuring Baryon Acoustic Oscillations \citep{2023arXiv230609407X}, Redshift Space Distortion \citep{2023ApJ...945L..30O} and primordial non-Gaussianity \citep{2023arXiv230202925K} using IA measurements. Moreover, IA of galaxies is highly related to their formation histories since the strength of IA varies with galaxy properties such as luminosity, stellar mass, color and morphology \citep{2015SSRv..193..139K,2015MNRAS.450.2195S,2022MNRAS.514.1021J,2022arXiv221211319S,2023arXiv230204230X}. Hence, a comprehensive understanding of IA is of great importance for a wide range of cosmological and astrophysical studies.

IA of dark matter (DM) halos is well described by the linear alignment (LA) model on large scales \citep{2001MNRAS.320L...7C,2004PhRvD..70f3526H,2011JCAP...05..010B,2019PhRvD.100j3506B,2020MNRAS.493L.124O}. The results are verified and extended to non-linear scales using the N-body simulations \citep{2000MNRAS.319..649H,2000ApJ...545..561C,2002MNRAS.335L..89J,2017ApJ...848...22X,2020MNRAS.494..694O}. However, IA of galaxies is much less well understood due to our limited knowledge of galaxy formation processes. Many studies found that IA of galaxies shows a complicated dependence on galaxy and their host halo properties \citep{2006MNRAS.369.1293Y,2015MNRAS.450.2195S,2016MNRAS.462.2668T,2016MNRAS.461.2702C,2020ApJ...904..135Y,2022MNRAS.514.1021J,2022PhRvD.106l3510H,2022arXiv221211319S}. To date, most of the IA studies focus on directly investigating the IA strength of galaxies, which is actually a combined effect of IA of halos and galaxy-halo alignment \citep{2020MNRAS.491.4116B}. Because IA of halos is well understood, studying the alignment between galaxies and their host halos would provide more valuable insights.

In observation, there were first studies that measure the alignment between galaxies and their host halos \citep{2009ApJ...694..214O,2009ApJ...694L..83O,2010MNRAS.402.2127S}. In these works, the misalignment angle between galaxies and their host halos are constrained  by fitting the galaxy-ellipticity (GI) or ellipticity-ellipticity (II) correlation functions with an assumed distribution function. In particular, assuming a Gaussian function with zero mean for the misalignment angle distribution, \citet{2009ApJ...694..214O} and \citet{2009ApJ...694L..83O} reported a dispersion of $35^{\circ}$ for the luminous red galaxies (LRG) from the Sloan Digital Sky Survey \citep[SDSS;][]{2000AJ....120.1579Y} DR6. \citet{2023arXiv230204230X} extended this method to SDSS-III Baryon Oscillation Spectroscopic Survey DR12 \citep[BOSS;][]{2015ApJS..219...12A,2016MNRAS.455.1553R} LOWZ and CMASS LRG samples with the better image data from the DESI Legacy Imaging Surveys \citep{2019AJ....157..168D}. They found that the alignment of central elliptical galaxies with their host halos increases monotonically with galaxy stellar mass or host halo mass, and that central elliptical galaxies are more aligned with their host halos when they evolve to a lower redshift. They also reported a weaker alignment between disk galaxies and their host halos. For comparison, \citet{2022PhRvD.106l3510H} modeled the 3D misalignment angles of the LOWZ and Dark Energy Survey \citep[DES;][]{2005astro.ph.10346T} samples assuming a Misis-Fisher distribution, and they found nearly no luminosity or redshift dependence for central galaxies.

In theory, the galaxy-halo alignment has been investigated in many works using the hydrodynamical simulations. Using the Horizon-AGN simulation \citep{2016MNRAS.463.3948D}, \citet{2017MNRAS.472.1163C} found that the misalignment angles of the minor axes of galaxies and their host halos decrease with halo mass and increase with redshift, and the shapes of galaxies and halos are not strongly related. \citet{2014MNRAS.441..470T} reported a similar halo mass dependent on major axis but with no redshift evolution in the MassiveBlack-II simulation \citep{2015MNRAS.450.1349K}. In EAGLE \citep{2015MNRAS.446..521S} and Cosmo-OWLS \citep{2010MNRAS.402.1536S} simulation, \citet{2015MNRAS.453..721V} also reported a halo mass dependence of the misalignment angle, but they found that early-type galaxies are more misaligned with their host halos compared to late-type galaxies, in contrast to the finding of \citet{2017MNRAS.472.1163C}. 
Furthermore, \citet{2015MNRAS.453..721V} found that the cosine of the 2D misalignment angle can be described by a double Gaussian plus a ‘floor’ with total 5 parameters and \citet{2014MNRAS.441..470T} provided a shifted exponential function with 3 parameters. Both studies reported that the assumption of a single Gaussian distribution may lead to an overestimation of the mean misalignment angles. However, it is hard to use either of the two models in the observation, since they have too many parameters. The only information from observation that can be used to constrain the misalignment is the suppression of the amplitudes of GI or II correlations, which is not sufficient to break the degeneracy of the parameters.

Therefore, in this work, on one hand, we seek to develop a model that can be utilized to constrain the galaxy-halo alignment in observation; and on the other hand, we aim to study the physical origin of the galaxy-halo alignment, which is still under debate. Using the TNG300-1 run from the IllustrisTNG simulations, We will construct a model with only \textit{one} free parameter to describe the 2D misalignment angle distributions, which can be applied to model the IA statistics in observations. We will also investigate the dependence of the misalignment angle on various galaxy/halo properties and endeavor to identify the most fundamental parameter that governs the galaxy-halo alignment. This paper is organized as follows. In Section \ref{sec:2}, we summarize the details of the simulation. We construct a model for 2D misalignment angles in Section \ref{sec:3} and investigate dependence of galaxy-halo alignment on galaxy or halo properties in Section \ref{sec:4}. We summarize our results in Section \ref{sec:5}.

\section{Data and Methods}\label{sec:2}
In this section, we provide a brief overview of the simulation used in this study, as well as the shape measurement method and galaxy/halo properties that we employ.

\subsection{Hydrodynamical simulations}
We use data from the IllustrisTNG simulations\footnote{www.tng-project.org} \citep{2018MNRAS.480.5113M, 2018MNRAS.475..624N, 2018MNRAS.477.1206N,2018MNRAS.475..676S,2018MNRAS.475..648P,2019ComAC...6....2N}, which is a suite of magnetohydrodynamic cosmological simulations. The simulations are run by the moving mesh code {\texttt{AREPO}} \citep{2010MNRAS.401..791S} and include various baryonic processes implemented as sub-grid physics \citep{2017MNRAS.465.3291W,2018MNRAS.473.4077P}. The cosmological parameters used in the simulations are consistent with Planck results \citep{2016A&A...594A..13P}: $\Omega_m=0.3089$, $\Omega_{\Lambda}=0.6911$, $\Omega_b=0.0486$, $h=0.6774$, $\sigma_8=0.8159$, $n_s=0.9667$. Among the simulation suite, we use the run with the largest box, TNG300-1 (hereafter TNG300). The box size is $205h^{-1}{\rm{Mpc}}$ with $2500^{3}$ DM particles and $2500^{3}$ gas cells, corresponding to mass resolutions of $m_{\rm{DM}}=5.9\times10^{7}h^{-1}M_{\odot}$ and $m_{\rm{gas}}=1.4\times10^{7}h^{-1}M_{\odot}$. The DM halos within the simulation were catalogued using friends-of-friends (FoF) methods \citep{1985ApJ...292..371D}, and the subhalos were catalogued using the {\texttt{SUBFIND}} algorithm \citep{2001MNRAS.328..726S}. The simulation output is stored in 100 snapshots for redshifts between $z=20$ and $z=0$. 

To study the evolution of galaxy-halo alignment, we choose the nearest snapshots around 5 reshifts $z=0.0,0.3,0.6,1.0,1.5$ in this work. To make sure galaxies are well resolved in shape, we only consider central galaxies with stellar mass $M_{*}>10^{10.0}h^{-1}M_{\odot}$ ($>1000$ particles), where $M_{*}$ is defined as the stellar mass within twice the stellar half mass radius. The host DM halos of these galaxies are also well resolved in shape, which have halo mass $M_{{\rm{vir}}}>10^{12.0}h^{-1}M_{\odot}$ at all redshifts, where $M_{{\rm{vir}}}$ is the viral mass of the halos \citep{1998ApJ...495...80B}.  

\begin{figure*}
    \plotone{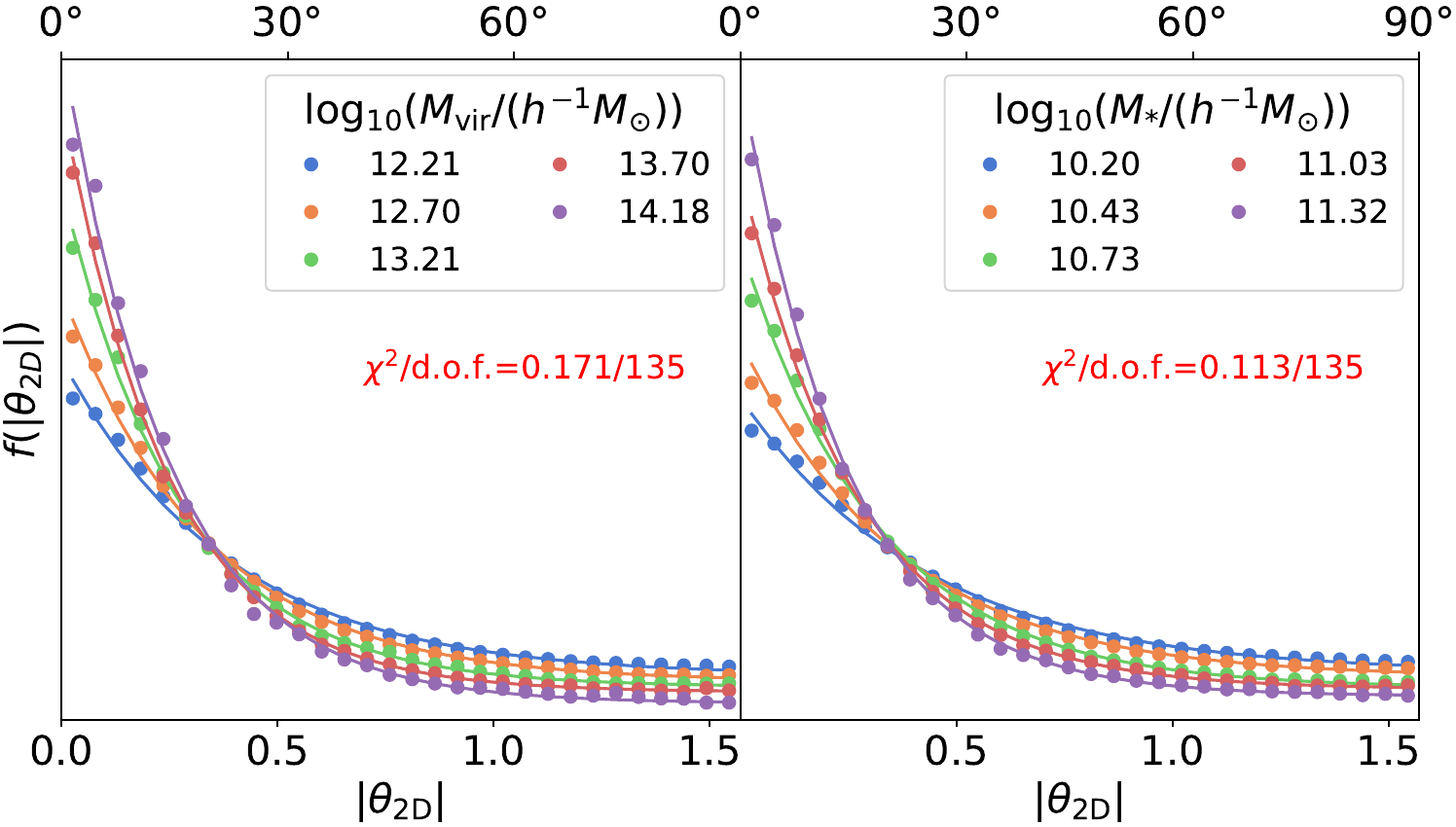}
    \caption{The distributions of $|\theta_{\rm{2D}}|$ for galaxies binned by $M_{\rm{vir}}$ (left) and $M_{*}$ (right) at $z=0$. Dots show the measurements from TNG300 and lines show the best-fit TSE distribution with {\textit{two}} free paramters $A$ and $B$.}
    \label{fig:2d_fit}
\end{figure*}

\begin{figure}
    \plotone{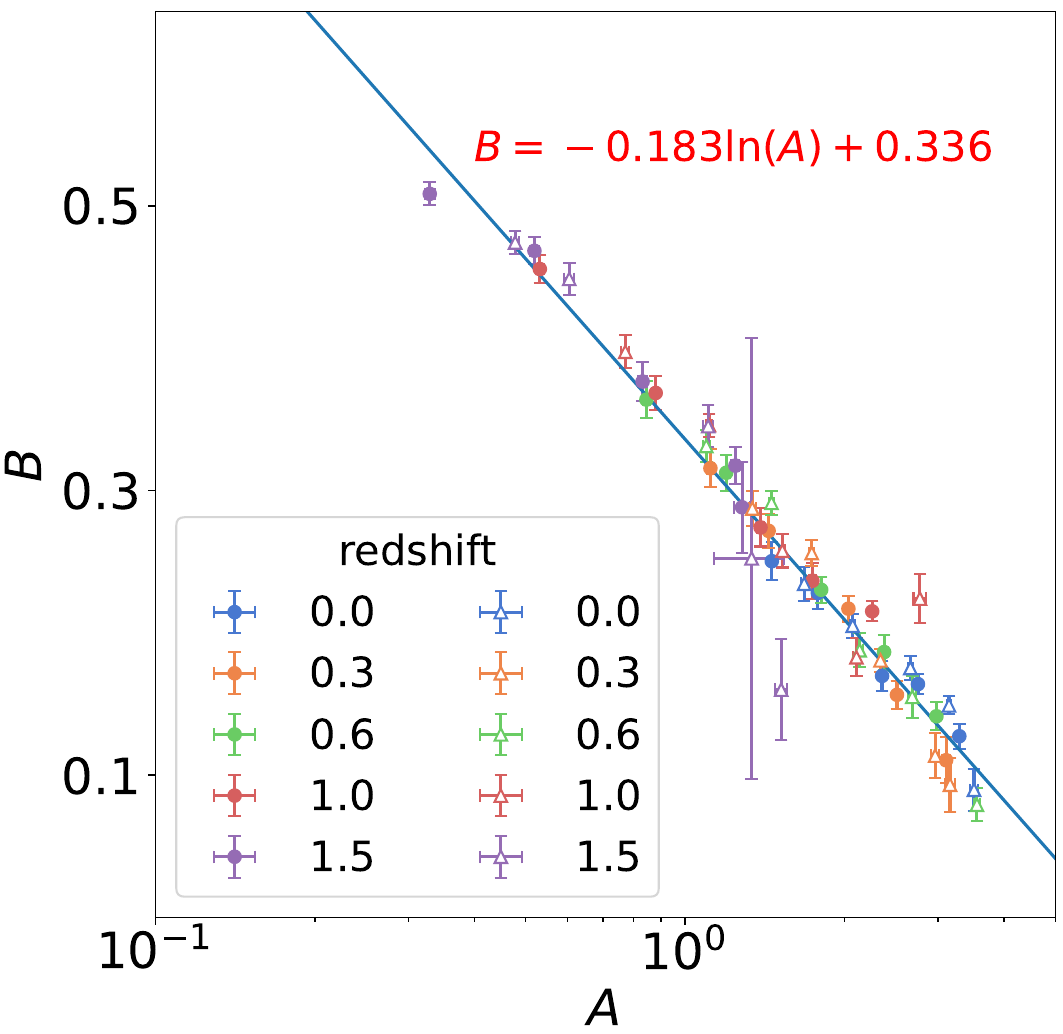}
    \caption{The best-fit parameters $A$ and $B$ of the TSE distribution for galaxies binned by $M_{*}$ (filled circles) and $M_{\rm{vir}}$ (open triangles) at various redshifts. The relation between $A$ and $B$ can be well described by a fitting formula $B=-0.183\ln(A)+0.336$ shown as the solid line.}
    \label{fig:2d_param}
\end{figure}

\subsection{Shapes of galaxies and halos}
We assume that the isodensity surfaces of both galaxies and halos can be described by triaxial ellipsoids \citep{2002ApJ...574..538J}. The principal axes and axial ratios of the galaxies and halos can be obtained by finding the eigenvectors and eigenvalues of their moment of inertia tensors. We use the reduced form of the inertia tensor \citep{2005ApJ...627..647B}
\begin{equation}
    I_{ij}^{{\rm{3D}}}=\sum_k\frac{x_{k,i}x_{k,j}}{r_k^2},\ {\rm{with}}\ i,j=\{1,2,3\}\,\,,
\end{equation}
where $x_{k,i}$ is the $i$-axis coordinate of the $k$th particle and $r_k$ is the distance of the $k$th particle to the galaxy/halo center, which is defined as the position of the particle with the minimum gravitational potential energy.

One problem in determining the shapes is how to select the star or DM particles used to calculate the inertia tensor. We use an iteration method to achieve it \citep{1991ApJ...368..325K,1995MNRAS.276..417J,2002MNRAS.335L..89J}. First, for galaxies, we select all the star particles within twice the stellar half mass radius, and for their DM halos, we select all the DM particles in the central subhalos. Here we use the central subhalos other than the whole halos to exclude the substructures, which may influence the stability of the iteration method. Then, we calculate $I_{i,j}$ and derive the three principal axes and minor-to-major ($c/a$) and meidan-to-major ($b/a$) axial ratios, where we define the lengths of the semi-major, median and minor axes as $a$, $b$ and $c$. Next, we re-calculate the new principal axes and axial ratios using the particles within an ellipsoid with the principal axes and axial ratios just determined. Here we set the major axis of the ellipsoid to $r_{98}$, which is defined as the distance of the particle with a distance larger than $98\%$ of the particles considered. We use $r_{98}$ rather than the maximum distance to exclude some extreme cases. We repeat the calculation for the updated ellipsoid until the axial ratios $c/a$ and $b/a$ converge to an accuracy of $1\%$. 

We also investigate the alignment of the 2D shapes of the projected galaxies and halos in this work. In this case, the iteration process is still done in 3D as above, and we use the converged particles to calculate the reduced 2D inertia tensor
\begin{equation}
I_{ij}^{{\rm{2D}}}=\sum_k\frac{x_{k,i}x_{k,j}}{r_{p,k}^2},\ {\rm{with}}\ i,j=\{1,2\}\,\,,
\end{equation}
where the inertia tensor is now reduced according to the projected distance $r_p$.

We have also examined our findings using the non-reduced inertia tensor. We observe that the misalignment angles, on the whole, are larger when employing this definition. Nevertheless, the relationship between misalignment angles and galaxy/halo properties remains unaltered, and our conclusions remain consistent.

\begin{figure*}
    \plotone{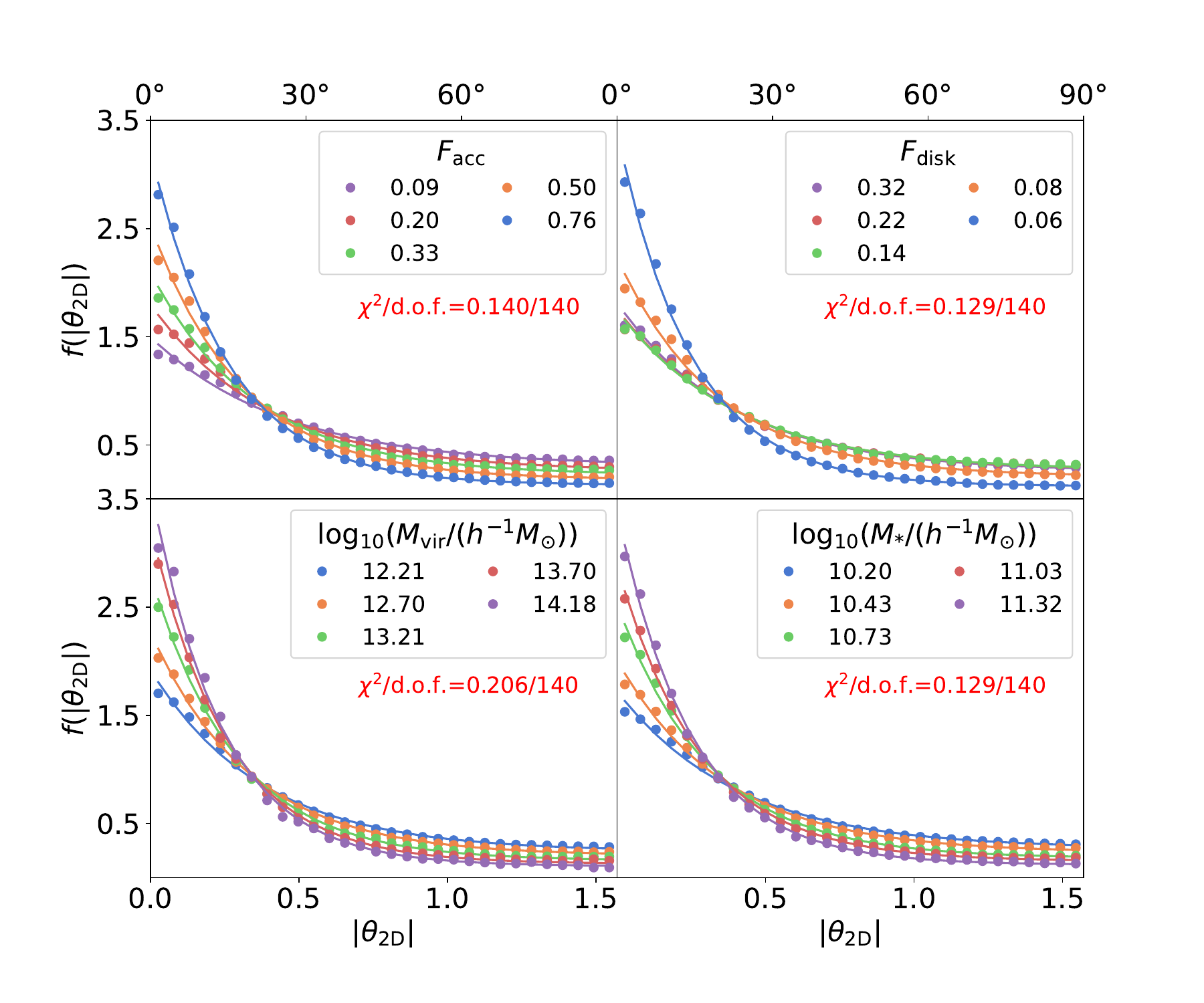}
    \caption{The distributions of $|\theta_{\rm{2D}}|$ for galaxies binned by different galaxy or halo properties at $z=0$. Dots show the measurements from TNG300 and lines show the best-fit TSE distribution with only {\textit{one}} free parameter $A$. For better illustration, we only display results from 5 out of the total 10 bins for $F_{\rm{acc}}$ and $F_{\rm{disk}}$.}
    \label{fig:2d_fit_one}
\end{figure*}

\subsection{List of other properties}
To better understand the physical origin of galaxy-halo alignment, we aim to study the dependence of galaxy-halo alignment on different galaxy or halo properties. we consider 4 specific properties in this work:
\begin{itemize}
    \item $F_{\rm{{acc}}}$: The amount of stellar mass that was formed ex situ, i.e., accrated from galaxy mergers. This quantity is provided in the stellar assembly catalog of TNG300 \citep{2015MNRAS.449...49R,2016MNRAS.458.2371R}.
    \item $F_{{\rm{disk}}}$: The disk-to-total stellar mass fraction. Disk particles are defined by $J_{z}/J(E)>0.7$, where $J_{z}$ is specific angular momentum of the particle and $J(E)$ is the expected angular momentum for a circular orbit at the same position as that particle. See \citet{2015ApJ...804L..40G} for more details of the definition.
    \item $M_{{\rm{vir}}}$: Viral mass of DM halos, defined as the halo mass within the viral radius $R_{\rm{vir}}$ derived using the fitting formula from \citet{1998ApJ...495...80B}.
    \item $M_{*}$: Stellar mass of central galaxies, defined as the stellar mass within twice the stellar half mass radius.
\end{itemize}

\section{Realistic modeling of the 2D galaxy-halo alignment}\label{sec:3}
In this section, we present a model for constraining the 2D galaxy-halo alignment in observations and compare it to the approaches employed in previous studies \citep{2009ApJ...694L..83O,2023arXiv230204230X}.

\subsection{Truncated shifted exponential (TSE) distribution}
There are several works that have attempted to model 2D misalignment angle $\theta_{\rm{2D}}$ between the 2D major axes of galaxies and their host DM halos using a Gaussian distribution with zero mean \citep{2009ApJ...694..214O,2009ApJ...694L..83O,2023arXiv230204230X}. However, it has been suggested by several studies that this assumption might not be necessarily true \citep{2014MNRAS.441..470T,2015MNRAS.453..721V,2017MNRAS.472.1163C}. In Figure \ref{fig:2d_fit}, we show distributions of $\theta_{\rm{2D}}$ for galaxies binned by $M_{\rm{vir}}$ or by $M_{*}$ at $z=0$. Equal logarithmic intervals of $0.5$ and $0.3$ are used for dividing the halos in the mass range of $[10^{12.0},10^{14.5}]h^{-1}M_{\odot}$ and the galaxies in the stellar mass range of $[10^{10.0},10^{11.5}]h^{-1}M_{\odot}$ respectively. For each galaxy, we measure $\theta_{\rm{2D}}$ in 1000 random projections. Since $\theta_{\rm{2D}}$ has a symmetrical distribution around $0$, we only consider its absolute value $|\theta_{\rm{2D}}|$. We find that the distributions of $|\theta_{\rm{2D}}|$ exhibit the same form when binned by either $M_{\rm{vir}}$ or $M_{*}$, and can be accurately described by a truncated shifted exponential (TSE) function, similar to the distribution reported by \citet{2014MNRAS.441..470T}
\begin{equation}
    f(|\theta_{\rm{2D}}|) = 
    \begin{cases}
    Ae^{-|\theta_{\rm{2D}}|/\tau}+B & |\theta_{\rm{2D}}|\leq\pi/2\\
    0 & {\rm{otherwise}}\,\,.
    \end{cases}\label{eq:ste}
\end{equation}
We derive the expectation of this distribution for convenience
\begin{equation}
\mathbb{E}[|\theta_{\rm{2D}}|]=A\tau\left(\tau - \frac{1}{2} e^{-\frac{\pi}{2\tau}} (2\tau + \pi)\right) + \frac{\pi^2 B}{8}\,\,, \label{eq:mean}
\end{equation}
and we also provide an algorithm for generating random numbers in Appendix \ref{sec:A}. 

Although there are three parameters, $A$, $B$, and $\tau$ in the TSE distribution, only two of them are free due to the normalization condition $\int_0^{\pi/2}f(|\theta_{\rm{2D}}|)d|\theta_{\rm{2D}}|=1$. Mathematically, it is more convenient to keep $\tau$ as a free parameter and express $A$ or $B$ with the other two. However, as we will demonstrate below, we aim to refine TSE to a distribution with a single free parameter by utilizing the empirical relations between the parameters. In this case, it is not recommended to treat $\tau$ as a free parameter since it becomes divergent when $A$ approaches $0$ and $B$ approaches $2/\pi$. This can make it challenging to obtain a reliable value for $\tau$. Therefore, we keep $A$ and $B$ as free parameters and express $\tau$ in terms of them
\begin{equation}
    Z = \frac{A\pi}{B\pi-2}\exp{(\frac{A\pi}{B\pi-2})}\,\,,
\end{equation}
\begin{equation}
    \tau = -\frac{\pi(\pi B-2)}{(4-2\pi B)W_0(Z)+2\pi A}\,\,,
\end{equation}
where $W_0$ refers to the principal branch of the Lambert $W$ function $W_n$\footnote{https://en.wikipedia.org/wiki/Lambert\_W\_function}, which can be accurately evaluated numerically\footnote{https://docs.scipy.org/doc/scipy/reference/generated/\\scipy.special.lambertw.html} \citep{corless1996lambert}. We show in Figure \ref{fig:2d_fit} the best-fit TSE distribution with 2 free parameters $A$ and $B$ for $|\theta_{\rm{2D}}|$. To quantify the goodness of fit, we define $\chi^2$ as
\begin{equation}
    \chi^2=\sum_i\frac{(f_i-f_i^{\rm{fit}})^2}{f_i^{\rm{fit}}}\,\,.
\end{equation}
Here, the summation is carried out over all the bins corresponding to $|\theta_{\text{2D}}|$ used in the fitting process. We find that TSE distribution fits the measurements very well for galaxies binned by $M_{*}$ or $M_{\rm{vir}}$.

\begin{figure}
    \plotone{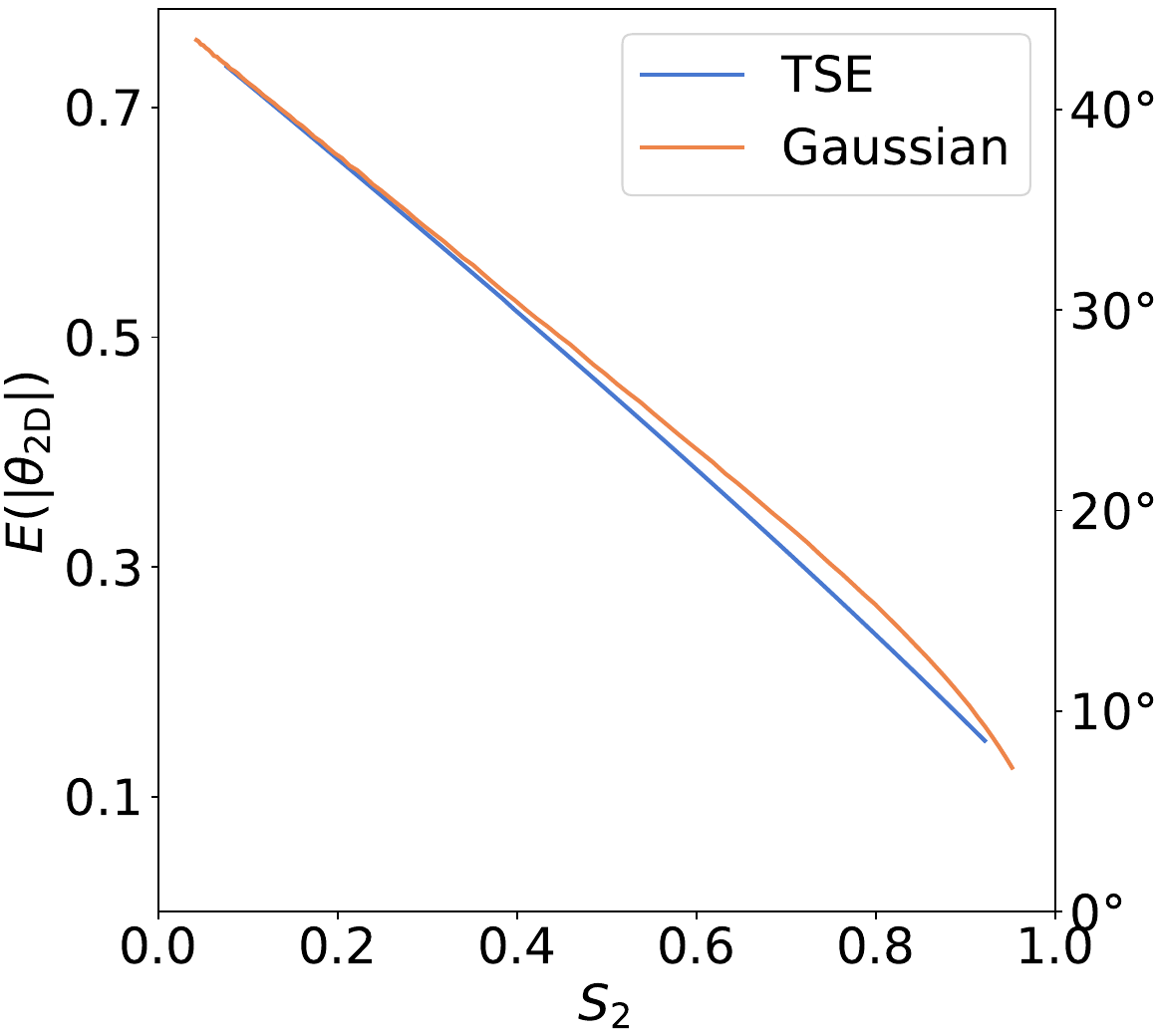}
    \caption{The relations between the suppression $S_2$ on $\tilde{w}_{g_+}$ and the mean $|\theta_{{\rm{2D}}}|$ from TSE and Gaussian distributions.}
    \label{fig:S}
\end{figure}

Currently, in observations, $\theta_{\rm{2D}}$ can only be constrained by analyzing the suppression of amplitude in GI or II correlations between galaxies and their corresponding dark matter halos \citep{2009ApJ...694..214O,2009ApJ...694L..83O,2023arXiv230204230X}. However, this information alone is insufficient to fully constrain the parameters as they can be highly degenerate. Therefore, we further reduce the degree of freedom for $A$ and $B$ using empirical relations. We obtain $A$ and $B$ by fitting the distribution of $|\theta_{\rm{2D}}|$ for galaxies binned by $M_{\rm{vir}}$ and $M_{*}$ at different redshifts. The results are presented in Figure \ref{fig:2d_param}. We observe a tight correlation between $A$ and $B$, which can be well described by a fitting formula of the form $B=-0.183\ln(A)+0.336$. Therefore, we refine the TSE distribution to have only one free parameter, which is useful for modeling $\theta_{\rm{2D}}$ in observations. The tight correlation between $A$ and $B$ suggests the possibility of a single process governing the galaxy-halo alignment, which will be explored in the next section.

To validate our TSE model, we perform additional fittings to $|\theta_{\rm{2D}}|$ distributions using only one parameter $A$ for galaxies binned by $F_{\rm{acc}}$, $M_{\rm{vir}}$, $F_{\rm{disk}}$, or $M_{*}$ at $z=0$. For $F_{\rm{acc}}$ and $F_{\rm{disk}}$, galaxies are divide into 10 equal sub-samples. The fitting results are displayed in Figure \ref{fig:2d_fit_one}. Remarkably, our model fits the measurement pretty well for all galaxy properties, and Figure \ref{fig:2d_param} indicates that this conclusion should hold for other redshifts as well. Moreover, we observe that the alignment of galaxies and their host DM halos increases with $F_{\rm{acc}}$, $M_{\rm{vir}}$, or $M_{*}$ increase, but decreases with $F_{\rm{disk}}$. These trends are in agreement with real observation reported by \citet{2023arXiv230204230X}, which will be explored in more detail in Section \ref{sec:4}.

It may not be immediately apparent that the TSE distribution should be able to describe the distributions of $|\theta_{\rm{2D}}|$ for galaxies with different properties. If we regard one property as a more fundamental factor in determining galaxy-halo alignment, the distribution of $|\theta_{\rm{2D}}|$ for galaxies binned by other properties can be viewed as a combination of TSE distributions determined by this property, which still following a TSE distribution. However, the sum of exponential functions is not mathematically guaranteed to be an exponential function. Nevertheless, in Appendix \ref{sec:B}, we show that for exponential functions with close exponents $\tau$, the sum of exponential functions can still be well described by an exponential function, validating the self-consistency of the TSE model.

\begin{figure*}
    \plotone{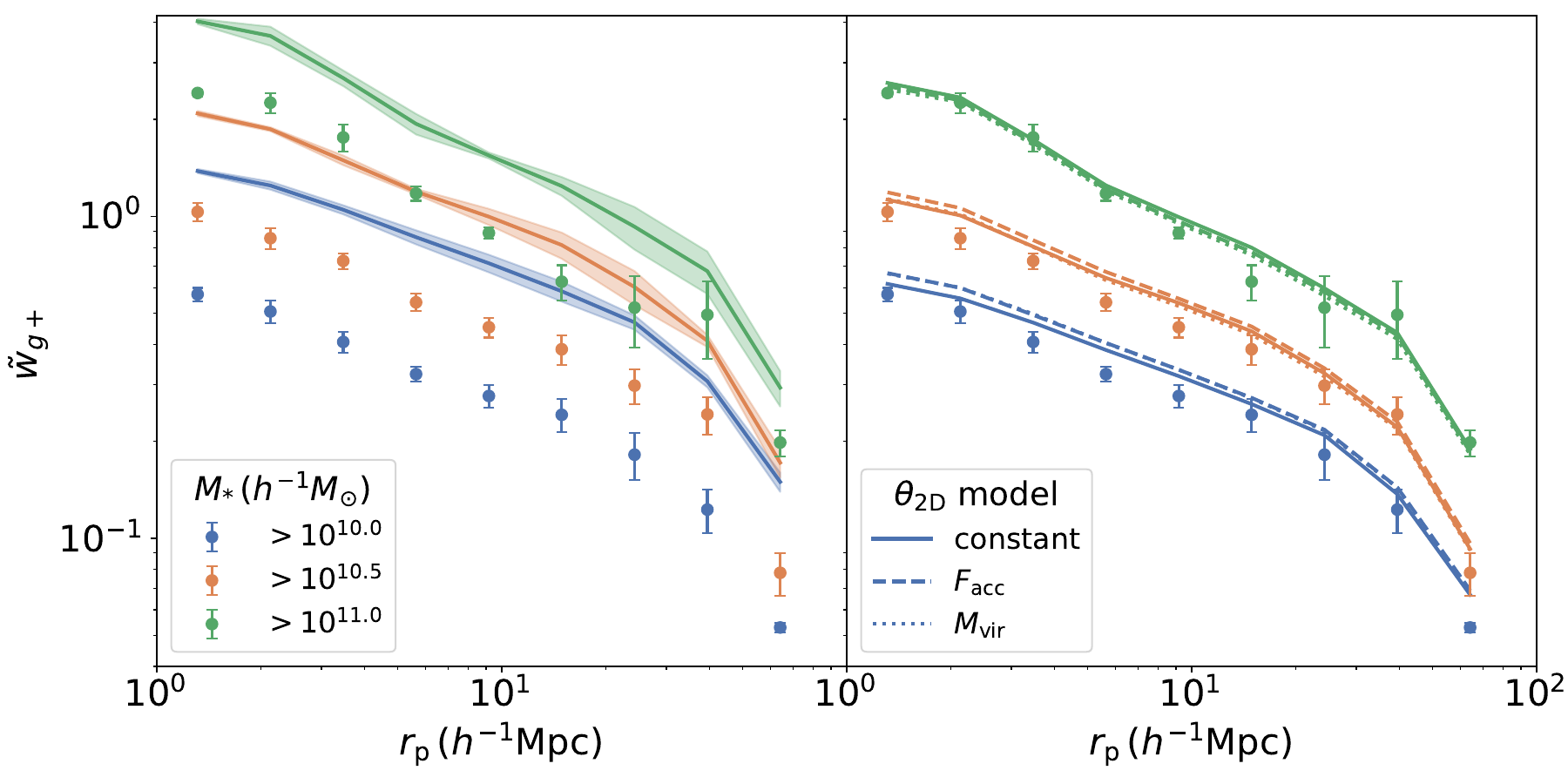}
    \caption{Left: The projected GI correlations $\tilde{w}_{g_+}$ of galaxies (dots) and halos (lines) for the 3 stellar mass limited samples at $z=0$. Right: $\tilde{w}_{g_+}$ of halos after suppressing the amplitudes with 2D misalignment angles from different models: a constant TSE model (solid lines), a $F_{\rm{acc}}$ dependent TSE model (dashed lines) and a $M_{\rm{vir}}$ dependent TSE model (dotted lines). The results are compared with $\tilde{w}_{g_+}$ of their galaxies (dots).}
    \label{fig:wg}
\end{figure*}

\subsection{Modeling galaxy-ellipticity (GI) correlation functions}
Our TSE distribution can be useful for modeling GI or II correlations in observation and constraining the misalignment in the real universe. This can be accomplished by examining the amplitude suppression in GI or II correlations due to the stochastic angular misalignment \citep{2009ApJ...694..214O,2009ApJ...694L..83O,2023arXiv230204230X}. 

In 2D, the shapes of galaxies and halos can be described by a two-component ellipticity, which is defined as
\begin{equation}
    e_{(+,\times)}=\frac{1-q^2}{1+q^2}(\cos{2\theta},\sin{2\theta})\,\,,\label{eq:1}
\end{equation}
where $q$ is the 2D minor-to-major axial ratio of the projected shape, and $\theta$ is the angle between the major axis projected on to the celestial sphere and the projected separation vector pointing to a specific object. As in \citet{2023arXiv230204230X}, we set $q=0$ for all galaxies and halos, and only care about their orientations. The $q=0$ GI correlation is defined as 
\begin{equation}
    \tilde{\xi}_{g+}(\bm{r})=\langle[1+\delta_{g_1}(\bm{x}_1)][1+\delta_{g_2}(\bm{x}_2)]e_+(\bm{x}_2)\rangle\,\,,
\end{equation}
where $\bm{r}=\bm{x}_1-\bm{x}_2$. This correlation can be estimated using the generalized Landy–Szalay estimator
 \citep{1993ApJ...412...64L,2006MNRAS.367..611M} with two random samples $R_s$ and $R$ corresponding to the tracers of ellipticity and density fields respectively,
\begin{equation}
    \tilde{\xi}_{g+}(r_{{\rm{p}}},\Pi) = \frac{S_{+}(D-R)}{R_{s}R}\,\,,
\end{equation}
where $R_sR$ is the normalized counts of random–random pairs in
a particular bin in the space of $(r_{{\rm{p}}},\Pi)$. $S_+D$ is the sum of the $+$ component of ellipticity in all pairs:
\begin{equation}
    S_+D=\sum_{i,j|r_{{\rm{p}}},\Pi}\frac{e_+(j|i)}{2\mathcal{R}}\,\,,
\end{equation}\label{eq:3}
where the ellipticity of the $j$th objects in the ellipticity tracers is defined relative to the direction to the $i$th objects in the density tracers, and $\mathcal{R}=1-\langle e_+^2 \rangle$ is the shape responsivity \citep{2002AJ....123..583B}. $\mathcal{R}$ equals to $0.5$ under our assumption of $q=0$. $S_+R$ is calculated in a similar way using the random catalog. The projected GI correlation are 
\begin{equation}
    \tilde{w}_{g_+}(r_{{\rm{p}}})=\int_{-\Pi_{{\rm{max}}}}^{\Pi_{{\rm{max}}}}\tilde{\xi}_{g+}(r_{{\rm{p}}},\Pi)d\Pi\,\,.
\end{equation}
We adopt $\Pi_{{\rm{max}}}=40\ h^{-1}{\rm{Mpc}}$ in the following.

\begin{figure*}
    \plotone{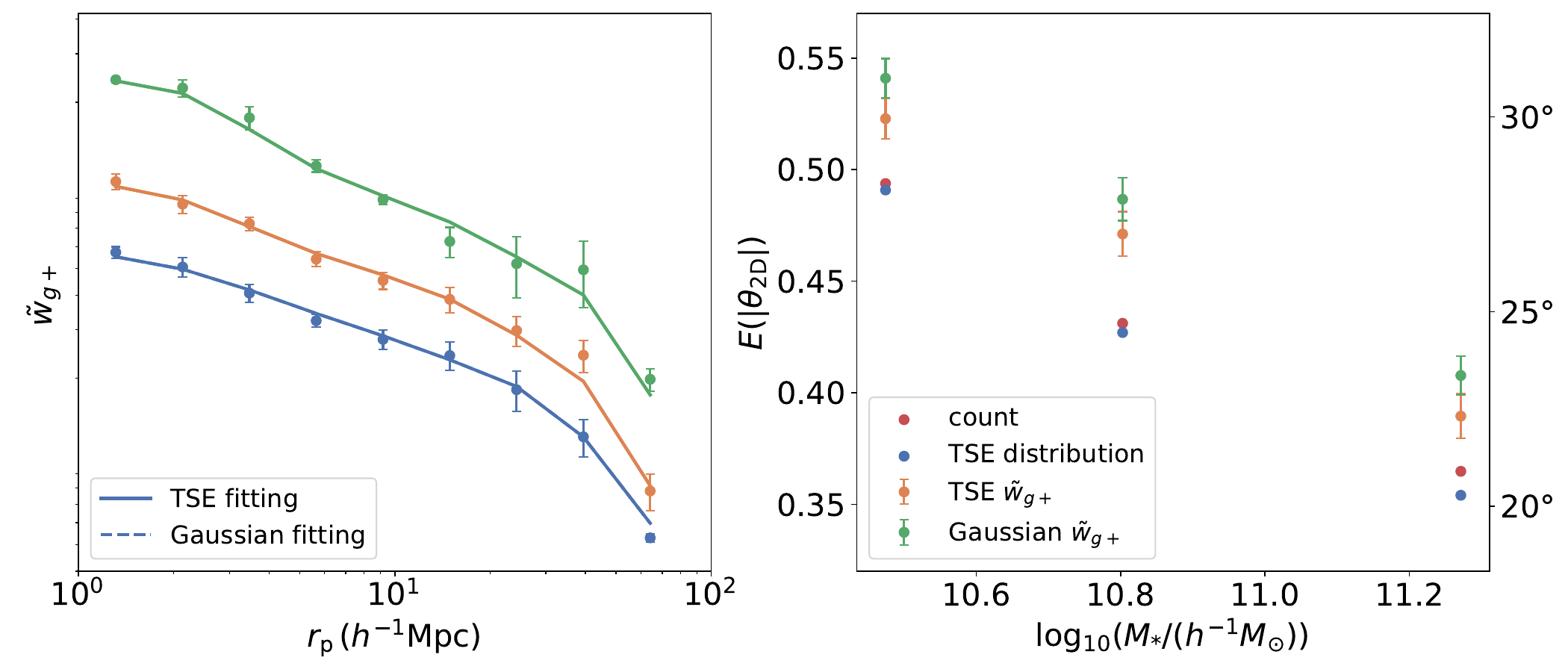}
    \caption{Left: Comparison between the galaxy and best-fit halo $\tilde{w}_{g_+}$ using both TSE and Gaussian misalignment models. Right: Mean $|\theta_{\rm{2D}}|$ obtained from different methods: direct counting (red), fitting the $|\theta_{\rm{2D}}|$ distribution with TSE model (blue), fitting $\tilde{w}_{g_+}$ with TSE model (orange), and fitting $\tilde{w}_{g_+}$ with Gaussian model (green).}
    \label{fig:wg_fitting}
\end{figure*}

To constrain $\theta_{{\rm{2D}}}$, previous studies typically involve recalculating $\tilde{w}_{g_+}$ for halos after perturbing the orientation of each halo according to the $\theta_{{\rm{2D}}}$ distribution. However, as demonstrated in \citet{2011JCAP...05..010B}, it is possible to analytically derive the suppression of amplitude $S_2$ in $\tilde{w}_{g_+}$ caused by a particular distribution of $\theta_{\rm{2D}}$. Therefore, an alternative approach is to directly constrain the suppression $S_2$ and then derive $\theta_{\rm{2D}}$, eliminating the need to recalculate $\tilde{w}_{g_+}$ using the estimator for each $\theta_{\rm{2D}}$ distribution. This method offers the advantages of being significantly faster and more stable. To establish the relation between $S_2$ and the distribution of $\theta_{\rm{2D}}$, we introduce the alignment correlation function, $w_g(r_{{\rm{p}}},\theta)$, which describes the dependence of clustering on both the projected separation, $r_{{\rm{p}}}$, and the orientation angle, $\theta$, measured from the axis of separation \citep{2009RAA.....9...41F}.
\begin{equation}
    \tilde{w}_{g+}(r_{{\rm{p}}})=\frac{2}{\pi}\int_0^{\pi/2}d\theta\cos{(2\theta)}w_g(r_{{\rm{p}}},\theta)
\end{equation}
If the misalignment angle $\theta_{\rm{2D}}$ is independent of $\theta$, then
\begin{align}
    \tilde{w}_{g+}^{\rm{mis}}(r_{{\rm{p}}})&=\frac{2}{\pi}\int_0^{\pi/2}d\theta\int_0^{\infty} f(\theta_{{\rm{2D}}})d\theta_{{\rm{2D}}}\cos{[2(\theta+\theta_{{\rm{2D}}})]}w_g(r_{{\rm{p}}},\theta)\notag \\
    &=\tilde{w}_{g+}(r_{{\rm{p}}})\int_0^{\infty}d\theta_{\rm{2D}}\cos{(2\theta_{\rm{2D}})}f(|\theta_{{\rm{2D}}}|)\notag \\
    &=S_2\tilde{w}_{g+}(r_{{\rm{p}}})\,\,,\\
    S_2&=\int_0^{\infty}d\theta_{\rm{2D}}\cos{(2\theta_{\rm{2D}})}f(|\theta_{{\rm{2D}}}|)\,\,.\label{eq:s_inter}
\end{align}
Hence, once $f(|\theta_{{\rm{2D}}}|)$ is known, the value of $S_2$ can be determined. For TSE, 
\begin{equation}
    S_2^{{\rm{TSE}}}=\frac{A\tau(e^{-\pi/(2\tau)}+1)}{4\tau^2+1}\,\,,\label{eg:S_TSE}
\end{equation}
similarly, for Gaussian distribution with zero mean and $\theta_{{\rm{2D}}}\in(-\infty,\infty)$,
\begin{equation}
    S_2^{{\rm{Gauss}}}=e^{-2\sigma^2}\,\,.
\end{equation}
The relations between $S_2$ and the mean $|\theta_{\rm{2D}}|$ for TSE and Gaussian distribution are shown in Figure \ref{fig:S}. In order to achieve the same suppression $S_2$, the Gaussian distribution requires a slightly larger mean value of $|\theta_{{\rm{2D}}}|$ than TSE, similar to that found in previous studies \citep{2014MNRAS.441..470T,2015MNRAS.453..721V}. We also confirm that the suppressed GI correlation recalculated using the estimator indeed yields the same value of $S_2$ as derived by Equation \ref{eq:s_inter}.

Figure \ref{fig:S} illustrates that it is always possible to find a Gaussian distribution and a TSE distribution that produce the same $S_2$ value, and this also holds true for any other distribution form of $\theta_{\text{2D}}$. Consequently, relying solely on $\tilde{w}_{g+}$ is insufficient to discern the distribution form of $\theta_{\text{2D}}$. It is important to note that different distribution forms of $\theta_{\text{2D}}$ can have distinct effects on $w_g(r_{\text{p}},\theta)$ and higher-order IA statistics. Here we investigate the effects on $w_g(r_{\rm{p}},\theta)$ as an example. Following \citet{2011JCAP...05..010B}, we can expend $w_g(r_{\rm{p}},\theta)$ as
\begin{equation}
    w_g(r_{\rm{p}},\theta)=w_g(r_{\rm{p}})+\sum_{n\in2\mathbb{Z}}a_n(r_{\rm{p}})\cos{(n\theta)}\,\,,
\end{equation}
with $a_2(r_{\rm{p}})=2\tilde{w}_{g+}(r_{\rm{p}})$. In the LA model, where only $\cos{(2\theta)}$ term is present, it is still possible to identify a Gaussian distribution and a TSE distribution of $\theta_{\text{2D}}$ that generate the same $w_g(r_{\text{p}},\theta)$. However, at nonlinear scale, where $n>2$ terms contribute to $w_g(r_{\rm{p}},\theta)$, the suppressed $w_g(r_{\rm{p}},\theta)$ is
\begin{equation}
    w_g^{\rm{mis}}(r_{\rm{p}},\theta)=w_g(r_{\rm{p}})+\sum_{n\in2\mathbb{Z}}S_na_n(r_{\rm{p}})\cos{(n\theta)}\,\,,
\end{equation}
\begin{equation}
    S_n=\int^{\infty}_0d\theta_{\rm{2D}}\cos{(n\theta_{\rm{2D}})}f(|\theta_{\rm{2D}}|)\,\,.
\end{equation}
For TSE,
\begin{align}
    S_n^{\rm{TSE}}=&\frac{Ae^{-\pi/(2\tau)}\tau(n\tau\sin{(\frac{\pi n}{2})}-\cos{(\frac{\pi n}{2})}+e^{\pi/(2\tau)})}{n^2\tau^2+1} \notag \\
    &+\frac{B\sin{(\frac{\pi n}{2})}}{n}\,\,.
\end{align}
For Gaussian,
\begin{equation}
    S_n^{\rm{Gauss}}=e^{-\frac{1}{2}n^2\sigma^2}\,\,.
\end{equation}
It is not possible to find a Gaussian distribution and a TSE distribution that yield the same value for all $S_n$, allowing us to distinguish the distribution form. However, angular misalignment significantly suppresses the nonlinear terms. As a result, measuring the $n>2$ terms for galaxies may be exceedingly difficult. Nevertheless, since $w_g(r_{\rm{p}},\theta)$ remains a two-point statistic, we expect the higher-order measurements possess additional information that can be used to constrain the distribution form.

\subsection{Testing the model in TNG300}
To test the validity of our method, we investigate whether the GI correlations of galaxies can be accurately reproduced by applying the best-fit $|\theta_{\rm{2D}}|$ distributions to their host halos. Due to the limited volume of TNG300, we are unable to study $\tilde{w}_{g_+}$ for small stellar mass bins as done in \citet{2023arXiv230204230X}. Therefore, we focus on 3 stellar mass limited samples as shape field tracers with $M_{*}>10^{10.0}h^{-1}M_{\odot}$, $10^{10.5}h^{-1}M_{\odot}$, and $10^{11.0}h^{-1}M_{\odot}$ at $z=0$. These samples have mean stellar masses of $10^{10.47}h^{-1}M_{\odot}$, $10^{10.80}h^{-1}M_{\odot}$, and $10^{11.27}h^{-1}M_{\odot}$, respectively. To improve the measurements, we use all halos with more than 50 particles at $z=0$ as the density field tracers for all three shape samples.

\begin{figure*}
    \plotone{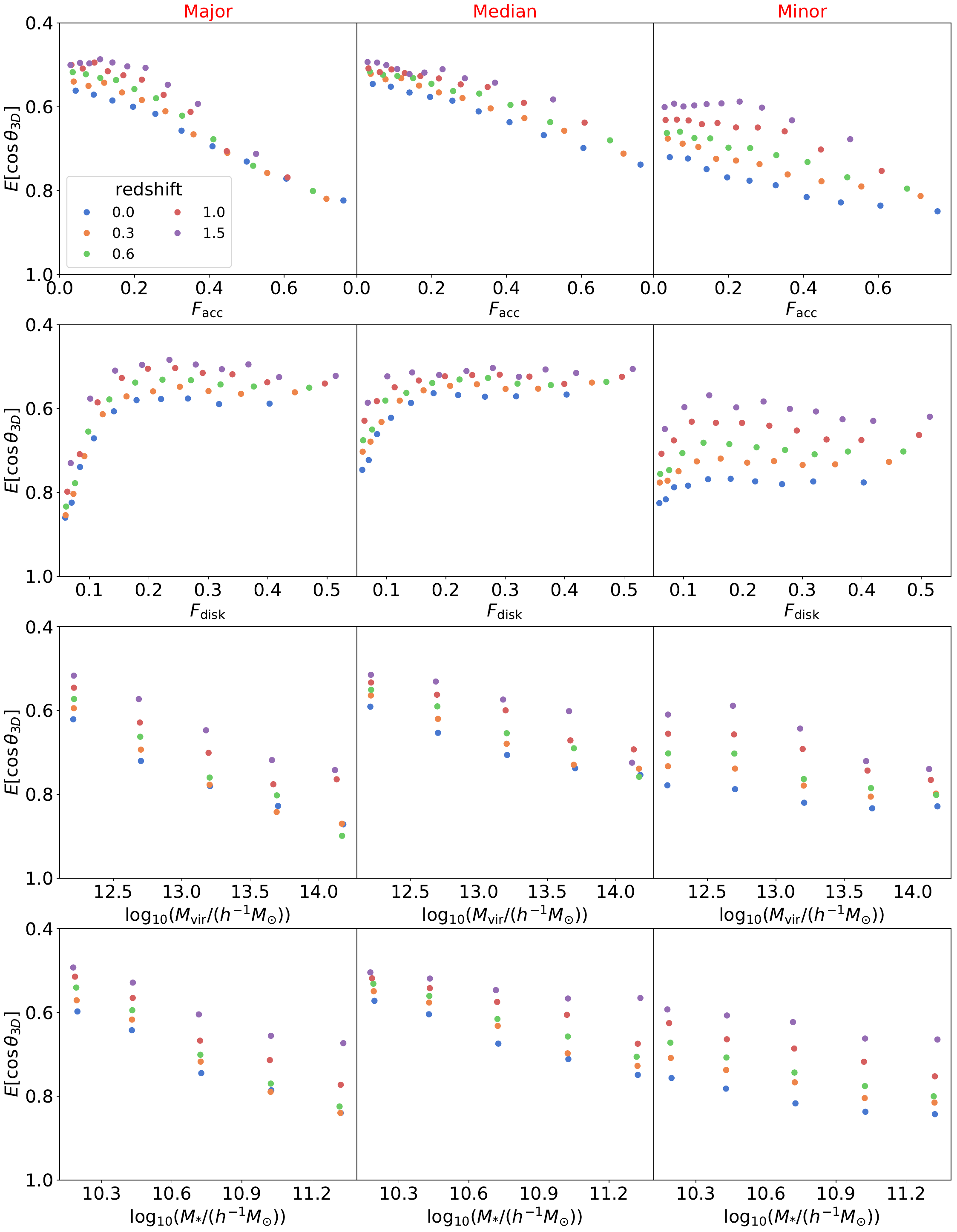}
    \caption{The mean $\cos{(\theta_{\rm{3D}})}$ for galaxies binned by different properties and at different redshifts. Each column shows the results for one principle axis and each row shows the results for the same property. Results from different redshifts are displayed by different colors. All the properties are plotted with the mean value in each bin.}
    \label{fig:3d_dep}
\end{figure*}

\begin{figure*}
    \plotone{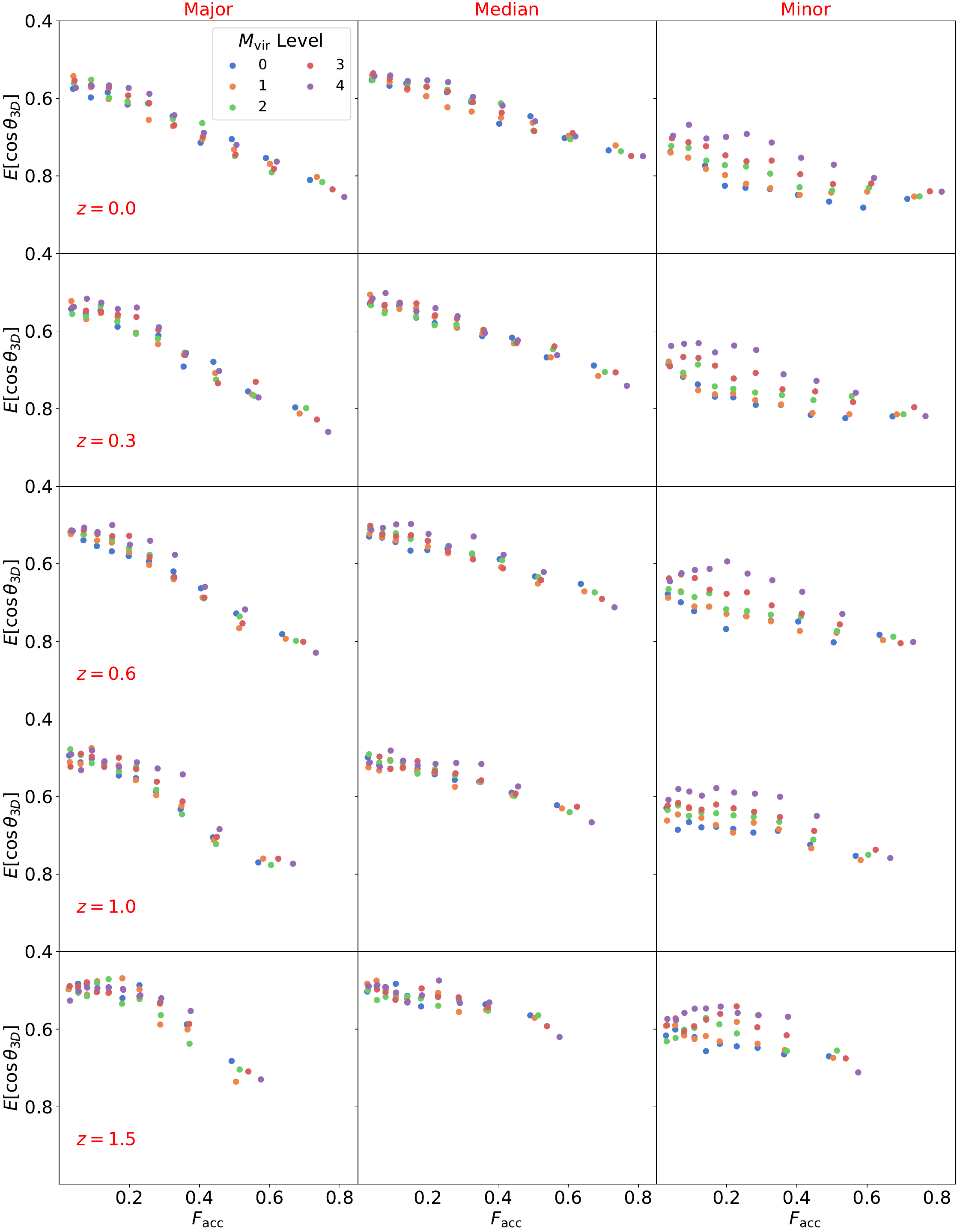}
    \caption{The mean $\cos{(\theta_{\rm{3D}})}$ for galaxies binned first by $F_{\rm{acc}}$ and further by $M_{\rm{vir}}$. Galaxies at each redshift are first divided into 10 equal sub-samples sorted by $F_{\rm{acc}}$. In each $F_{\rm{acc}}$ bin, they are further divided into 5 equal sub-samples sorted by $M_{\rm{vir}}$. Different $M_{\rm{vir}}$ bins in each $F_{\rm{acc}}$ bin are plotted with different colors, with $0$ indicating the lowest $M_{\rm{vir}}$ and $4$ indicating the highest. Each column shows the results for one principle axis and each row shows the results for the same redshift. All results are plotted with the mean $F_{\rm{acc}}$ in each bin.}
    \label{fig:3d_facc}
\end{figure*}

\begin{figure*}
    \plotone{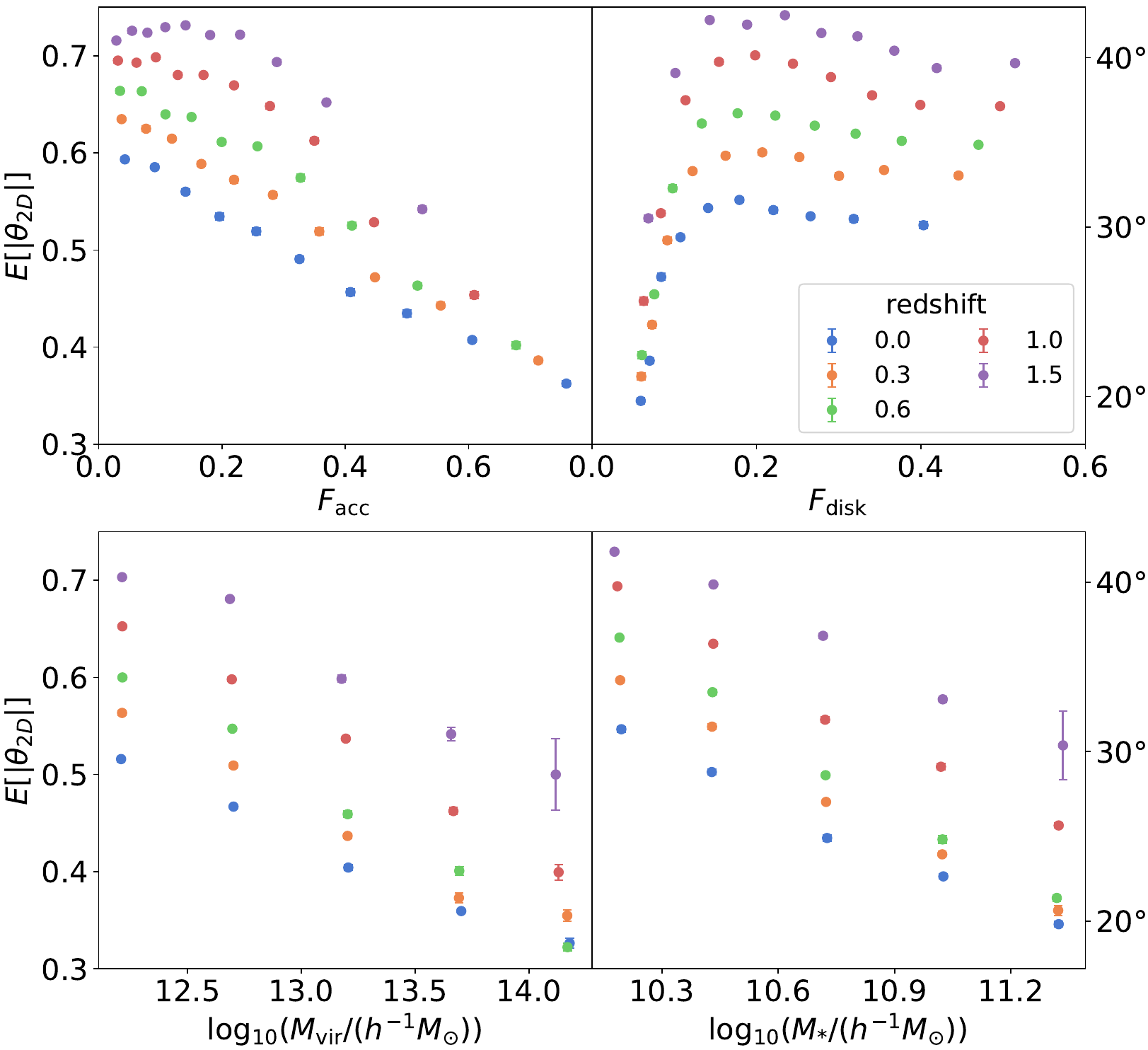}
    \caption{The mean $|\theta_{\rm{2D}}|$ of galaxies binned by different properties and at different redshifts. Results from different redshifts are displayed by different colors. All the properties are plotted with the mean value in each bin.}
    \label{fig:2d_dep}
\end{figure*}

\begin{figure*}
    \plotone{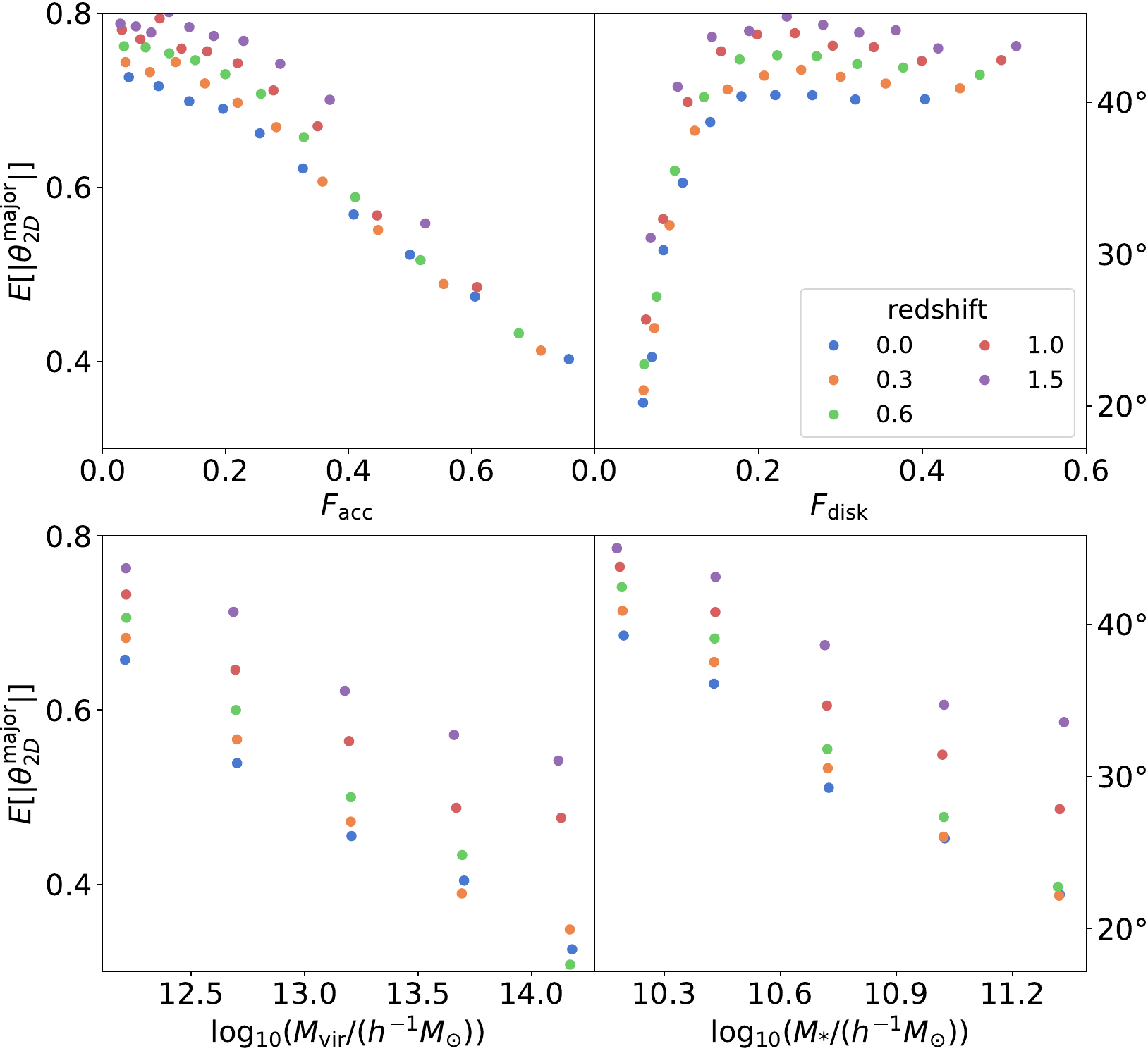}
    \caption{The mean $|\theta_{\rm{2D}}^{\rm{major}}|$ from the projection of only 3D major axes rather than all particles. Galaxies are binned by different properties and at different redshifts. Results from different redshifts are displayed by different colors. All the properties are plotted with the mean value in each bin.}
    \label{fig:2d_dep_major}
\end{figure*}

\begin{figure*}
    \plotone{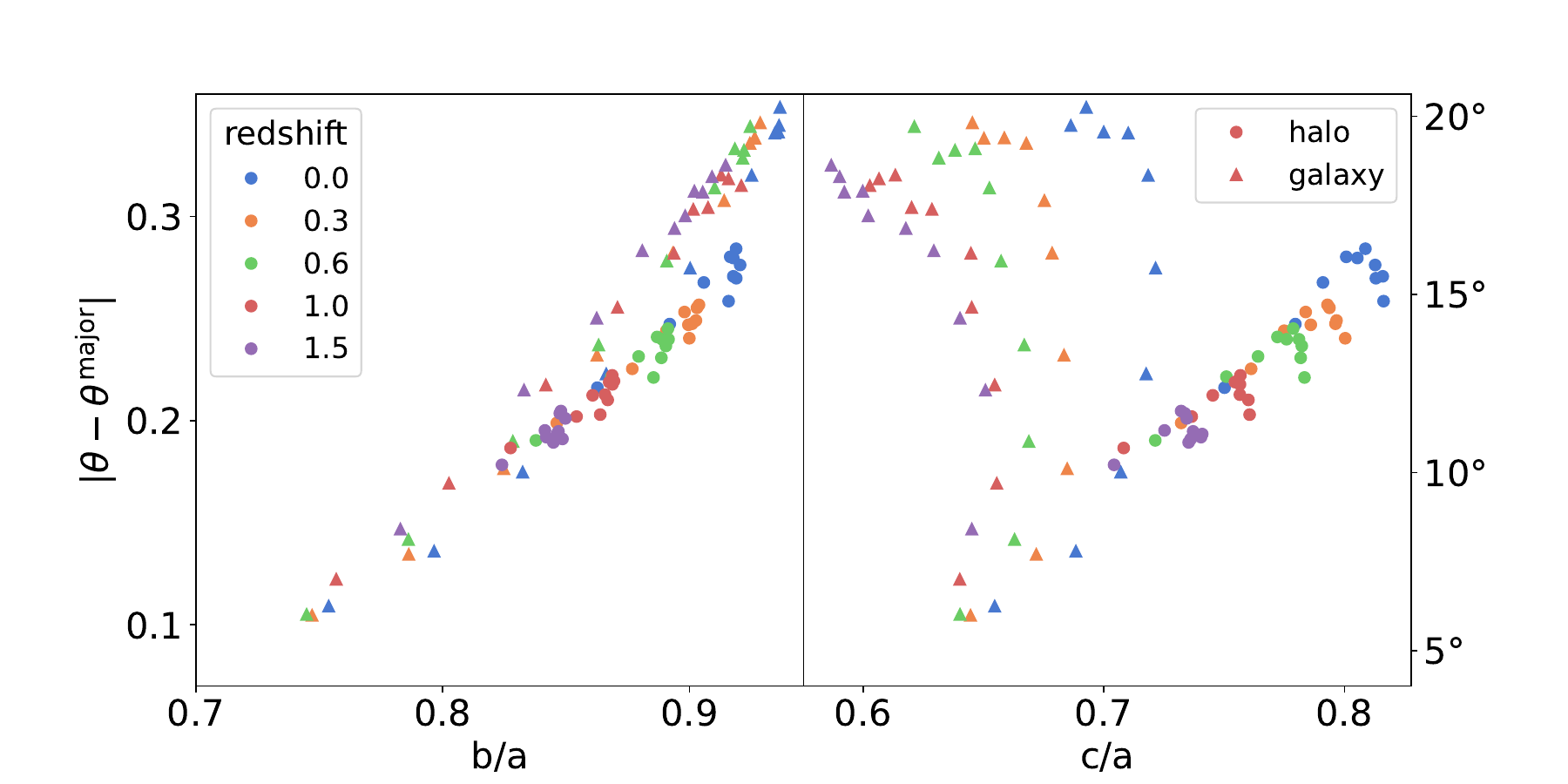}
    \caption{Difference between the position angles of 2D major axes obtained by projecting all particles $\theta$ and only 3D major axes $\theta^{\rm{major}}$ of galaxies and halos, as a function of mean axis ratios $b/a$ (left) and $c/a$ (right). Galaxies and halos are binned by $F_{\rm{acc}}$ at each redshift. Results from different redshifts are displayed by different colors, with dots for halos and triangles for galaxies.}
    \label{fig:2d_delta}
\end{figure*}

We present the $\tilde{w}_{g_+}$ measurements of galaxies and their host halos in the left panel of Figure \ref{fig:wg}. Mean values and errors of $\tilde{w}_{g_+}$ are estimated from three different projection directions. Our results indicate that galaxies have lower $\tilde{w}_{g_+}$ than their host halos, which is expected due to the misalignment. Furthermore, we observe that more massive galaxies and their host halos exhibit a larger $\tilde{w}_{g_+}$, consistent with previous studies \citep{2002MNRAS.335L..89J,2017ApJ...848...22X}. 

Subsequently, we explore whether suppressing the amplitude of $\tilde{w}_{g+}$ of host halos, based on the TSE distributions of $|\theta_{\rm{2D}}|$, can accurately reproduce the $\tilde{w}_{g+}$ of their corresponding galaxies. We consider three different models: a constant TSE model, a $F_{\rm{acc}}$-dependent TSE model, and a $M_{\rm{vir}}$-dependent TSE model. In the constant TSE model, the suppression is calculated by the TSE distributions obtained by fitting the $|\theta_{\rm{2D}}|$ of the entire sample, which has $\mathbb{E}(|\theta_{\rm{2D}}|)=0.49$, $0.43$, and $0.35$ for $M_{*}>10^{10.0}h^{-1}M_{\odot}$, $10^{10.5}h^{-1}M_{\odot}$, and $10^{11.0}h^{-1}M_{\odot}$, respectively. For $F_{\rm{acc}}$- and $M_{\rm{vir}}$-dependent models, we use the best-fit relations based on the results in Figure \ref{fig:2d_fit_one}: $\mathbb{E}(|\theta_{{\rm{2D}}}|)=-0.33F_{{\rm{acc}}}+0.60$ and $\mathbb{E}(|\theta_{{\rm{2D}}}|)=-0.11M_{{\rm{vir}}}+1.83$. For the constant TSE model, the suppression $S_2$ is calculated using Equation \ref{eg:S_TSE}. In the $F_{\rm{acc}}$- and $M_{\rm{vir}}$-dependent models, we recalculate the correlations using the estimator after perturbing the orientation of each halo, although, in principle, they can also be derived analytically.

The results are shown in the right panel of Figure \ref{fig:wg}. We find that the suppressed GI correlations of halos by all three misalignment models can basically match the GI correlations of galaxies, but the amplitudes are slightly higher. The results from the three different models are consistent with each other, indicating that the GI correlations are not sensitive to the variation of misalignment angle distributions between halos and are mainly determined by the overall distribution. Despite the good agreement, the small residuals suggest that there may exist unknown couplings between the misalignment angles and other factors, such as the coupling with galaxy or halo shapes as reported by \citet{2009ApJ...694L..83O} and \citet{2011JCAP...05..010B}. Nevertheless, as we will show below, this model is already accurate, with only a $3^{\circ}$ bias for the worst-case. We leave the detailed investigation of the mismatch to future works.

We further constrain the misalignment angles of the three stellar mass limited samples by fitting their $\tilde{w}_{g_+}$ using the constant TSE model and compare them to the results obtained from directly counting and fitting the $|\theta_{{\rm{2D}}}|$ distribution. We also include the fitting results using a Gaussian model with zero mean as used in previous works \citep{2009ApJ...694..214O,2009ApJ...694L..83O,2023arXiv230204230X} for comparison.  The results are presented in Figure \ref{fig:wg_fitting}. The left panel shows that both the TSE and Gaussian models provide a good fit to $\tilde{w}_{g_+}$ of galaxies, and in the right panel, we observe that the Gaussian model requires slightly larger values ($\sim1^{\circ}$) of mean $|\theta_{\rm{2D}}|$ than the TSE model to achieve the same suppression for fitting $\tilde{w}_{g_+}$ of galaxies as expected from Figure \ref{fig:S}. We again demonstrate that the TSE model provides a good fit to the $|\theta_{\rm{2D}}|$ distribution, as shown by comparing the fitting results to the direct counting results. The results from fitting $\tilde{w}_{g_+}$ with the TSE model are slightly higher ($<3^{\circ}$) than those obtained from direct counting and fitting the distribution, which is consistent with the results in Figure \ref{fig:wg}. 

In summary, our TSE model reproduces the galaxy GI correlations well, with only a small bias, and has the potential to be useful for modeling the galaxy-halo alignment in observational studies.

\section{Dependence on Galaxy/Halo Properties}\label{sec:4}
Recently, in observation, \citet{2023arXiv230204230X} discovered that galaxies with larger stellar mass $M_{*}$ or host halo mass $M_{\rm{vir}}$ and at lower redshifts are more aligned with their host halos, though they focus mainly on massive galaxies ($M_{*}>10^{11.0}M_{\odot}$). They also observe that elliptical galaxies are considerably more aligned with their halos compared to disk galaxies. These trends are already evident in Figure \ref{fig:2d_fit_one} in TNG300. In this section, we aim to further investigate the dependence of galaxy-halo alignment on various galaxy or halo properties, both in 2D and 3D.

\subsection{3D misalignment angle}
We first investigate the 3D misalignment angle $\theta_{\rm{3D}}$ for all three principal axes. Since we haven't found a model that accurately describes the distribution of $\theta_{\rm{3D}}$, we directly calculate the mean value of $\cos{(\theta_{\rm{3D}})}$ for each property and at each redshift. We use $\cos{(\theta_{\rm{3D}})}$ instead of $\theta_{\rm{3D}}$ because $\theta_{\rm{3D}}$ occupies different differential volumes.

In Figure \ref{fig:3d_dep}, we show the mean $\cos{(\theta_{\rm{3D}})}$ of the three principle axes for galaxies with different properties and at different redshifts. We find that galaxies are more aligned with their host halos for larger $F_{\rm{acc}}$, $M_{\rm{vir}}$, $M_{*}$ and smaller $F_{\rm{disk}}$ for all three principle axes, although the mean $\cos{(\theta_{\rm{3D}})}$ seems to remain constant for $F_{\rm{disk}}>0.2$. Galaxies at lower redshifts are also more aligned with their host halos. An important question that arises is which parameter is more closely related to the origin of the alignment between galaxies and their host halos. To answer this question, we want to point out that the redshift evolution of the mean $\cos(\theta_{\rm{3D}})$ is much weaker when galaxies are binned by $F_{\rm{acc}}$ than when they are binned by $M_{\rm{vir}}$ and $M_{*}$, as shown in Figure \ref{fig:3d_dep}. Specifically, for the major axis, when binned by $F_{\rm{acc}}$, the redshift dependence is nearly eliminated for $F_{\rm{acc}}>0.5$. Although similar results are also found for $F_{\rm{disk}}<0.1$, there is no dependence of $\theta_{\rm{3D}}$ on $F_{\rm{disk}}$ for $F_{\rm{disk}}>0.2$. Moreover, theoretically, the alignment between galaxies and their host halos is expected to be influenced by their formation history. Galaxies formed mainly by mergers should have more similar formation history as their host halos. Therefore, we think $F_{\rm{acc}}$ might be a more fundamental property related to galaxy-halo alignment.

To prove this hypothesis, we investigate whether the dependence of $\theta_{\rm{3D}}$ on other parameters can be eliminated when controlling $F_{\rm{acc}}$. Hence, after dividing the galaxies into 10 equal sub-samples according to $F_{\rm{acc}}$, we further divide the galaxies into 5 equal sub-samples sorted by $M_{\rm{vir}}$ in each $F_{\rm{acc}}$ bins. We show the mean $\cos{(\theta_{\rm{3D}})}$ of each $M_{\rm{vir}}$ sub-samples in each $F_{\rm{acc}}$ bin in Figure \ref{fig:3d_facc}. We find that after binning by $F_{\rm{acc}}$, the major and median axes show almost no dependence on $M_{\rm{vir}}$ for all redshifts. However, for the minor axis, there is still a weak dependence on $M_{\rm{vir}}$, but in the opposite direction to the trend shown in Figure \ref{fig:3d_dep} where larger $M_{\rm{vir}}$ galaxies exhibit smaller mean $\cos(\theta_{\rm{3D}})$. This opposite dependence also appears to be present for the major and median axes, but to a much lesser extent.

The results presented in Figure \ref{fig:3d_dep} and Figure \ref{fig:3d_facc} suggest that $F_{\rm{acc}}$ is a crucial parameter for characterizing the galaxy-halo alignment. However, additional dependences of mean $\cos(\theta_{\rm{3D}})$ on redshift or $M_{\rm{vir}}$ after controlling for $F_{\rm{acc}}$ suggests the possibility of a secondary parameter affecting the alignment. Specifically, as demonstrated in Figure \ref{fig:3d_dep} and Figure \ref{fig:3d_facc}, after binned by $F_{\rm{acc}}$, the dependence of $\theta_{\rm{3D}}$ on redshift and $M_{\rm{vir}}$ is nearly negligible for galaxies formed predominantly through mergers ($F_{\rm{acc}}>0.5$). In contrast, galaxies formed primarily in situ ($F_{\rm{acc}}<0.5$) exhibit a clear dependence on redshift and $M_{\rm{vir}}$. This dependence may arise from the differences in the star formation and feedback processes experienced by galaxies at different redshifts and with varying host halo masses, which can potentially impact the galaxy-halo alignment. Additionally, we should exercise caution regarding the definition of $F_{\text{acc}}$, as it may not be equitable across different redshifts and galaxies. This discrepancy can lead to non-universal behavior. Further investigation is required to fully elucidate these effects. 

\subsection{2D misalignment angle}
We follow a similar approach as in 3D to study the dependence of 2D misalignment on different galaxy or halo properties. In Figure \ref{fig:2d_dep}, we show the mean $|\theta_{\rm{2D}}|$ obtained from the best-fit TSE distributions. We find similar trends as in 3D, where the mean $|\theta_{\rm{2D}}|$ decreases with $F_{\rm{acc}}$, $M_{\rm{vir}}$, and $M_{*}$, and increases with $F_{\rm{disk}}$ and redshift. However, we observe that the scatter between different redshifts is much larger compared to that in 3D. Specifically, for $F_{\rm{acc}}$, although $|\theta_{\rm{2D}}|$ also shows weak dependence on redshift at $F_{\rm{acc}}>0.5$, the scatter at $F_{\rm{acc}}<0.5$ is larger than in 3D. 

The larger difference in $|\theta_{\rm{2D}}|$ between different redshifts can be attributed to the evolution of galaxy and halo shapes. The orientation of the 2D major axis is determined by all three 3D principle axes, and the contribution of each 3D axis depends on the axial ratios $c/a$ and $b/a$. Therefore, even if $\theta_{\rm{3D}}$ remains constant with redshift, changes in axial ratios with redshift can cause $\theta_{\rm{2D}}$ to show dependence on redshift, as it contains different contributions from the three principle axes with different $\theta_{\rm{3D}}$. We verify the above conjecture by recalculating the 2D misalignment angles $|\theta_{\rm{2D}}^{\rm{major}}|$, where the 2D major axes of galaxies and halos are obtained from the projection of 3D major axes, which is equivalent to assuming that a galaxy or halo is a line along its major axis with $c/a=0$ and $b/a=0$. In this case, the shape evolution is eliminated. The results are shown in \ref{fig:2d_dep_major}. Comparing to $|\theta_{\rm{2D}}|$, we find that the difference in mean $|\theta_{\rm{2D}}^{\rm{major}}|$ between different redshifts is reduced, and the results are consistent with that from the mean $\cos(\theta_{\rm{3D}})$ of the 3D major axis. 

We further calculate the difference $|\theta-\theta^{\rm{major}}|$ between the orientations of the 2D major axes obtained by projecting all particles $\theta$ and only the 3D major axes $\theta^{\rm{major}}$, and analyze their dependence on axial ratios. We calculate the mean $|\theta-\theta^{\rm{major}}|$ and axis ratios $b/a$ and $c/a$ for galaxies and halos binned by $F_{\rm{acc}}$. As shown in Figure \ref{fig:2d_delta}, $|\theta-\theta^{\rm{major}}|$ increases monotonously with $b/a$ for both galaxies and halos. However, the slopes are slightly different for galaxies and halos, which may be due to their different distributions of $c/b$ at fixed $b/a$. We observe a similar relationship between $|\theta-\theta^{\rm{major}}|$ and $c/a$ for halos, while galaxies exhibit a different dependence. This indicates that there is a strong relationship with $b/a$ and $c/a$ for halos, while galaxies can have a more complex distribution of shapes. The tight relationship between $|\theta-\theta^{\rm{major}}|$ and $b/a$ suggests that the evolution of $b/a$ is the primary factor driving the larger differences in mean $|\theta_{\rm{2D}}|$ between different redshifts, with the behavior of $c/a$ acting as a secondary factor.

Therefore, although we observe similar dependence in 2D, we believe that studying the galaxy-halo alignment in 3D is much more informative.

\section{conclusion and discussion}\label{sec:5}
In this study, we investigate the alignment between galaxies and their host halos in both 3D and 2D using the IllustrisTNG 300-1 simulation. Our goal is to propose a model that can be used to constrain alignment of galaxies and their host halos in observational studies and understand the physical origin of the galaxy-halo alignment. Our main findings are as follows:
\begin{itemize}
    \item The distribution of $|\theta_{\rm{2D}}|$ can be well described by a truncated shifted exponential (TSE) distribution across different galaxy and halo properties and at different redshifts.
    \item With the TSE distribution, we can reproduce the galaxy GI correlation $\tilde{w}_{g+}$ from halos with only a small bias ($<3^{\circ}$). This bias may come from the unknown coupling between $\theta_{\rm{2D}}$ and other factors such as halo shapes.
    \item The 3D misalignment angles $\theta_{\rm{3D}}$ for all three principal axes and the 2D misalignment angles $\theta_{\rm{2D}}$ decrease with $F_{\rm{acc}}$, $M_{\rm{vir}}$, and $M_{*}$, and increase with $F_{\rm{disk}}$ and redshift, similar to those found by \citet{2023arXiv230204230X} in observation. 
    \item $F_{\rm{acc}}$ appears to be a more fundamental parameter that determines the galaxy-halo alignment than other parameters we studied. Grouping galaxies by $F_{\rm{acc}}$, the $M_{\rm{vir}}$ dependence of $\theta_{\rm{3D}}$ for all three principle axes is either eliminated or inverted. Moreover, the redshift dependence is also reduced comparing to other properties.
    \item The redshift dependence of $\theta_{\rm{2D}}$ is much larger than $\theta_{\rm{3D}}$, which may be due to the evolution of the 3D axis ratios $b/a$ and $c/a$ of both galaxies and halos. In detail, $\theta_{\rm{3D}}$ of the three principle axes contribute differently to $\theta_{\rm{2D}}$ at different redshift since galaxies and halos have different shape distributions.  
\end{itemize}

Although we have made significant progress in understanding the galaxy-halo alignment, there are still some remaining questions that require further investigation. Firstly, it would be interesting to investigate the origin of the remaining dependence on redshift and other properties of $\theta_{\rm{3D}}$ after controlling $F_{\rm{acc}}$. Secondly, it is important to eliminate the small bias in modeling the galaxy GI correlation. Thirdly, our study only considers the orientations of galaxies and halos. Further research is needed to explore the galaxy-halo shape relation and correlations between shapes and orientations to fully understand the intrinsic alignment. Lastly, it is essential to investigate the robustness of our results to the galaxy formation models used in the hydrodynamic simulations, and to test our conclusions with other simulations.

Furthermore, by combining the measurements of GI correlations \citep{2023arXiv230204230X}, the galaxy-halo connection \citep{2023ApJ...944..200X}, the halo intrinsic alignment model \citep{2017ApJ...848...22X}, and the galaxy-halo alignment model developed in this study, it becomes possible to model the intrinsic alignment of central galaxies using a halo model approach. 

Our results in this work can enhance the understanding of the galaxy-halo alignment and have important implications for intrinsic alignment (IA) correction in weak lensing studies, IA cosmology, and the theory of galaxy formation.

\section*{Acknowledgments}
The work is supported by NSFC (12133006, 11890691, 11621303), grant No. CMS-CSST-2021-A03, and 111 project No. B20019. We gratefully acknowledge the support of the Key Laboratory for Particle Physics, Astrophysics and Cosmology, Ministry of Education. This work made use of the Gravity Supercomputer at the Department of Astronomy, Shanghai Jiao Tong University.

%% For this sample we use BibTeX plus aasjournals.bst to generate the
%% the bibliography. The sample63.bib file was populated from ADS. To
%% get the citations to show in the compiled file do the following:
%%
%% pdflatex sample63.tex
%% bibtext sample63
%% pdflatex sample63.tex
%% pdflatex sample63.tex

\bibliography{sample63}{}

\begin{thebibliography}{}
\expandafter\ifx\csname natexlab\endcsname\relax\def\natexlab#1{#1}\fi
\providecommand{\url}[1]{\href{#1}{#1}}
\providecommand{\dodoi}[1]{doi:~\href{http://doi.org/#1}{\nolinkurl{#1}}}
\providecommand{\doeprint}[1]{\href{http://ascl.net/#1}{\nolinkurl{http://ascl.net/#1}}}
\providecommand{\doarXiv}[1]{\href{https://arxiv.org/abs/#1}{\nolinkurl{https://arxiv.org/abs/#1}}}

\bibitem[{{Alam} {et~al.}(2015){Alam}, {Albareti}, {Allende Prieto}, {Anders},
  {Anderson}, {Anderton}, {Andrews}, {Armengaud}, {Aubourg}, {Bailey}, {Basu},
  {Bautista}, {Beaton}, {Beers}, {Bender}, {Berlind}, {Beutler}, {Bhardwaj},
  {Bird}, {Bizyaev}, {Blake}, {Blanton}, {Blomqvist}, {Bochanski}, {Bolton},
  {Bovy}, {Shelden Bradley}, {Brandt}, {Brauer}, {Brinkmann}, {Brown},
  {Brownstein}, {Burden}, {Burtin}, {Busca}, {Cai}, {Capozzi}, {Carnero
  Rosell}, {Carr}, {Carrera}, {Chambers}, {Chaplin}, {Chen}, {Chiappini},
  {Chojnowski}, {Chuang}, {Clerc}, {Comparat}, {Covey}, {Croft}, {Cuesta},
  {Cunha}, {da Costa}, {Da Rio}, {Davenport}, {Dawson}, {De Lee}, {Delubac},
  {Deshpande}, {Dhital}, {Dutra-Ferreira}, {Dwelly}, {Ealet}, {Ebelke},
  {Edmondson}, {Eisenstein}, {Ellsworth}, {Elsworth}, {Epstein}, {Eracleous},
  {Escoffier}, {Esposito}, {Evans}, {Fan}, {Fern{\'a}ndez-Alvar}, {Feuillet},
  {Filiz Ak}, {Finley}, {Finoguenov}, {Flaherty}, {Fleming}, {Font-Ribera},
  {Foster}, {Frinchaboy}, {Galbraith-Frew}, {Garc{\'\i}a},
  {Garc{\'\i}a-Hern{\'a}ndez}, {Garc{\'\i}a P{\'e}rez}, {Gaulme}, {Ge},
  {G{\'e}nova-Santos}, {Georgakakis}, {Ghezzi}, {Gillespie}, {Girardi},
  {Goddard}, {Gontcho}, {Gonz{\'a}lez Hern{\'a}ndez}, {Grebel}, {Green},
  {Grieb}, {Grieves}, {Gunn}, {Guo}, {Harding}, {Hasselquist}, {Hawley},
  {Hayden}, {Hearty}, {Hekker}, {Ho}, {Hogg}, {Holley-Bockelmann}, {Holtzman},
  {Honscheid}, {Huber}, {Huehnerhoff}, {Ivans}, {Jiang}, {Johnson},
  {Kinemuchi}, {Kirkby}, {Kitaura}, {Klaene}, {Knapp}, {Kneib}, {Koenig},
  {Lam}, {Lan}, {Lang}, {Laurent}, {Le Goff}, {Leauthaud}, {Lee}, {Lee},
  {Licquia}, {Liu}, {Long}, {L{\'o}pez-Corredoira}, {Lorenzo-Oliveira},
  {Lucatello}, {Lundgren}, {Lupton}, {Mack}, {Mahadevan}, {Maia}, {Majewski},
  {Malanushenko}, {Malanushenko}, {Manchado}, {Manera}, {Mao}, {Maraston},
  {Marchwinski}, {Margala}, {Martell}, {Martig}, {Masters}, {Mathur},
  {McBride}, {McGehee}, {McGreer}, {McMahon}, {M{\'e}nard}, {Menzel},
  {Merloni}, {M{\'e}sz{\'a}ros}, {Miller}, {Miralda-Escud{\'e}}, {Miyatake},
  {Montero-Dorta}, {More}, {Morganson}, {Morice-Atkinson}, {Morrison},
  {Mosser}, {Muna}, {Myers}, {Nandra}, {Newman}, {Neyrinck}, {Nguyen},
  {Nichol}, {Nidever}, {Noterdaeme}, {Nuza}, {O'Connell}, {O'Connell},
  {O'Connell}, {Ogando}, {Olmstead}, {Oravetz}, {Oravetz}, {Osumi}, {Owen},
  {Padgett}, {Padmanabhan}, {Paegert}, {Palanque-Delabrouille}, {Pan},
  {Parejko}, {P{\^a}ris}, {Park}, {Pattarakijwanich}, {Pellejero-Ibanez},
  {Pepper}, {Percival}, {P{\'e}rez-Fournon}, {P{\'e}rez-R{\`a}fols},
  {Petitjean}, {Pieri}, {Pinsonneault}, {Porto de Mello}, {Prada}, {Prakash},
  {Price-Whelan}, {Protopapas}, {Raddick}, {Rahman}, {Reid}, {Rich}, {Rix},
  {Robin}, {Rockosi}, {Rodrigues}, {Rodr{\'\i}guez-Torres}, {Roe}, {Ross},
  {Ross}, {Rossi}, {Ruan}, {Rubi{\~n}o-Mart{\'\i}n}, {Rykoff},
  {Salazar-Albornoz}, {Salvato}, {Samushia}, {S{\'a}nchez}, {Santiago},
  {Sayres}, {Schiavon}, {Schlegel}, {Schmidt}, {Schneider}, {Schultheis},
  {Schwope}, {Sc{\'o}ccola}, {Scott}, {Sellgren}, {Seo}, {Serenelli}, {Shane},
  {Shen}, {Shetrone}, {Shu}, {Silva Aguirre}, {Sivarani}, {Skrutskie},
  {Slosar}, {Smith}, {Sobreira}, {Souto}, {Stassun}, {Steinmetz}, {Stello},
  {Strauss}, {Streblyanska}, {Suzuki}, {Swanson}, {Tan}, {Tayar}, {Terrien},
  {Thakar}, {Thomas}, {Thomas}, {Thompson}, {Tinker}, {Tojeiro}, {Troup},
  {Vargas-Maga{\~n}a}, {Vazquez}, {Verde}, {Viel}, {Vogt}, {Wake}, {Wang},
  {Weaver}, {Weinberg}, {Weiner}, {White}, {Wilson}, {Wisniewski},
  {Wood-Vasey}, {Ye`che}, {York}, {Zakamska}, {Zamora}, {Zasowski}, {Zehavi},
  {Zhao}, {Zheng}, {Zhou}, {Zhou}, {Zou}, \& {Zhu}}]{2015ApJS..219...12A}
{Alam}, S., {Albareti}, F.~D., {Allende Prieto}, C., {et~al.} 2015, \apjs, 219,
  12, \dodoi{10.1088/0067-0049/219/1/12}

\bibitem[{{Bailin} \& {Steinmetz}(2005)}]{2005ApJ...627..647B}
{Bailin}, J., \& {Steinmetz}, M. 2005, \apj, 627, 647, \dodoi{10.1086/430397}

\bibitem[{{Bernstein} \& {Jarvis}(2002)}]{2002AJ....123..583B}
{Bernstein}, G.~M., \& {Jarvis}, M. 2002, \aj, 123, 583, \dodoi{10.1086/338085}

\bibitem[{{Bhowmick} {et~al.}(2020){Bhowmick}, {Chen}, {Tenneti}, {Di Matteo},
  \& {Mandelbaum}}]{2020MNRAS.491.4116B}
{Bhowmick}, A.~K., {Chen}, Y., {Tenneti}, A., {Di Matteo}, T., \& {Mandelbaum},
  R. 2020, \mnras, 491, 4116, \dodoi{10.1093/mnras/stz3240}

\bibitem[{{Blazek} {et~al.}(2011){Blazek}, {McQuinn}, \&
  {Seljak}}]{2011JCAP...05..010B}
{Blazek}, J., {McQuinn}, M., \& {Seljak}, U. 2011, \jcap, 2011, 010,
  \dodoi{10.1088/1475-7516/2011/05/010}

\bibitem[{{Blazek} {et~al.}(2019){Blazek}, {MacCrann}, {Troxel}, \&
  {Fang}}]{2019PhRvD.100j3506B}
{Blazek}, J.~A., {MacCrann}, N., {Troxel}, M.~A., \& {Fang}, X. 2019, \prd,
  100, 103506, \dodoi{10.1103/PhysRevD.100.103506}

\bibitem[{{Brown} {et~al.}(2002){Brown}, {Taylor}, {Hambly}, \&
  {Dye}}]{2002MNRAS.333..501B}
{Brown}, M.~L., {Taylor}, A.~N., {Hambly}, N.~C., \& {Dye}, S. 2002, \mnras,
  333, 501, \dodoi{10.1046/j.1365-8711.2002.05354.x}

\bibitem[{{Bryan} \& {Norman}(1998)}]{1998ApJ...495...80B}
{Bryan}, G.~L., \& {Norman}, M.~L. 1998, \apj, 495, 80, \dodoi{10.1086/305262}

\bibitem[{{Catelan} {et~al.}(2001){Catelan}, {Kamionkowski}, \&
  {Blandford}}]{2001MNRAS.320L...7C}
{Catelan}, P., {Kamionkowski}, M., \& {Blandford}, R.~D. 2001, \mnras, 320, L7,
  \dodoi{10.1046/j.1365-8711.2001.04105.x}

\bibitem[{{Chisari} {et~al.}(2016{\natexlab{a}}){Chisari}, {Laigle}, {Codis},
  {Dubois}, {Devriendt}, {Miller}, {Benabed}, {Slyz}, {Gavazzi}, \&
  {Pichon}}]{2016MNRAS.461.2702C}
{Chisari}, N., {Laigle}, C., {Codis}, S., {et~al.} 2016{\natexlab{a}}, \mnras,
  461, 2702, \dodoi{10.1093/mnras/stw1409}

\bibitem[{{Chisari} \& {Dvorkin}(2013)}]{2013JCAP...12..029C}
{Chisari}, N.~E., \& {Dvorkin}, C. 2013, \jcap, 2013, 029,
  \dodoi{10.1088/1475-7516/2013/12/029}

\bibitem[{{Chisari} {et~al.}(2016{\natexlab{b}}){Chisari}, {Dvorkin},
  {Schmidt}, \& {Spergel}}]{2016PhRvD..94l3507C}
{Chisari}, N.~E., {Dvorkin}, C., {Schmidt}, F., \& {Spergel}, D.~N.
  2016{\natexlab{b}}, \prd, 94, 123507, \dodoi{10.1103/PhysRevD.94.123507}

\bibitem[{{Chisari} {et~al.}(2017){Chisari}, {Koukoufilippas}, {Jindal},
  {Peirani}, {Beckmann}, {Codis}, {Devriendt}, {Miller}, {Dubois}, {Laigle},
  {Slyz}, \& {Pichon}}]{2017MNRAS.472.1163C}
{Chisari}, N.~E., {Koukoufilippas}, N., {Jindal}, A., {et~al.} 2017, \mnras,
  472, 1163, \dodoi{10.1093/mnras/stx1998}

\bibitem[{Corless {et~al.}(1996)Corless, Gonnet, Hare, Jeffrey, \&
  Knuth}]{corless1996lambert}
Corless, R.~M., Gonnet, G.~H., Hare, D.~E., Jeffrey, D.~J., \& Knuth, D.~E.
  1996, Advances in Computational mathematics, 5, 329,
  \dodoi{10.1007/BF02124750}

\bibitem[{{Croft} \& {Metzler}(2000)}]{2000ApJ...545..561C}
{Croft}, R. A.~C., \& {Metzler}, C.~A. 2000, \apj, 545, 561,
  \dodoi{10.1086/317856}

\bibitem[{{Davis} {et~al.}(1985){Davis}, {Efstathiou}, {Frenk}, \&
  {White}}]{1985ApJ...292..371D}
{Davis}, M., {Efstathiou}, G., {Frenk}, C.~S., \& {White}, S.~D.~M. 1985, \apj,
  292, 371, \dodoi{10.1086/163168}

\bibitem[{{Dey} {et~al.}(2019){Dey}, {Schlegel}, {Lang}, {Blum}, {Burleigh},
  {Fan}, {Findlay}, {Finkbeiner}, {Herrera}, {Juneau}, {Landriau}, {Levi},
  {McGreer}, {Meisner}, {Myers}, {Moustakas}, {Nugent}, {Patej}, {Schlafly},
  {Walker}, {Valdes}, {Weaver}, {Y{\`e}che}, {Zou}, {Zhou}, {Abareshi},
  {Abbott}, {Abolfathi}, {Aguilera}, {Alam}, {Allen}, {Alvarez}, {Annis},
  {Ansarinejad}, {Aubert}, {Beechert}, {Bell}, {BenZvi}, {Beutler}, {Bielby},
  {Bolton}, {Brice{\~n}o}, {Buckley-Geer}, {Butler}, {Calamida}, {Carlberg},
  {Carter}, {Casas}, {Castander}, {Choi}, {Comparat}, {Cukanovaite}, {Delubac},
  {DeVries}, {Dey}, {Dhungana}, {Dickinson}, {Ding}, {Donaldson}, {Duan},
  {Duckworth}, {Eftekharzadeh}, {Eisenstein}, {Etourneau}, {Fagrelius},
  {Farihi}, {Fitzpatrick}, {Font-Ribera}, {Fulmer}, {G{\"a}nsicke},
  {Gaztanaga}, {George}, {Gerdes}, {Gontcho}, {Gorgoni}, {Green}, {Guy},
  {Harmer}, {Hernandez}, {Honscheid}, {Huang}, {James}, {Jannuzi}, {Jiang},
  {Joyce}, {Karcher}, {Karkar}, {Kehoe}, {Kneib}, {Kueter-Young}, {Lan},
  {Lauer}, {Le Guillou}, {Le Van Suu}, {Lee}, {Lesser}, {Perreault Levasseur},
  {Li}, {Mann}, {Marshall}, {Mart{\'\i}nez-V{\'a}zquez}, {Martini}, {du Mas des
  Bourboux}, {McManus}, {Meier}, {M{\'e}nard}, {Metcalfe},
  {Mu{\~n}oz-Guti{\'e}rrez}, {Najita}, {Napier}, {Narayan}, {Newman}, {Nie},
  {Nord}, {Norman}, {Olsen}, {Paat}, {Palanque-Delabrouille}, {Peng},
  {Poppett}, {Poremba}, {Prakash}, {Rabinowitz}, {Raichoor}, {Rezaie},
  {Robertson}, {Roe}, {Ross}, {Ross}, {Rudnick}, {Safonova}, {Saha},
  {S{\'a}nchez}, {Savary}, {Schweiker}, {Scott}, {Seo}, {Shan}, {Silva},
  {Slepian}, {Soto}, {Sprayberry}, {Staten}, {Stillman}, {Stupak}, {Summers},
  {Sien Tie}, {Tirado}, {Vargas-Maga{\~n}a}, {Vivas}, {Wechsler}, {Williams},
  {Yang}, {Yang}, {Yapici}, {Zaritsky}, {Zenteno}, {Zhang}, {Zhang}, {Zhou}, \&
  {Zhou}}]{2019AJ....157..168D}
{Dey}, A., {Schlegel}, D.~J., {Lang}, D., {et~al.} 2019, \aj, 157, 168,
  \dodoi{10.3847/1538-3881/ab089d}

\bibitem[{{Dubois} {et~al.}(2016){Dubois}, {Peirani}, {Pichon}, {Devriendt},
  {Gavazzi}, {Welker}, \& {Volonteri}}]{2016MNRAS.463.3948D}
{Dubois}, Y., {Peirani}, S., {Pichon}, C., {et~al.} 2016, \mnras, 463, 3948,
  \dodoi{10.1093/mnras/stw2265}

\bibitem[{{Faltenbacher} {et~al.}(2009){Faltenbacher}, {Li}, {White}, {Jing},
  {Shu-DeMao}, \& {Wang}}]{2009RAA.....9...41F}
{Faltenbacher}, A., {Li}, C., {White}, S. D.~M., {et~al.} 2009, Research in
  Astronomy and Astrophysics, 9, 41, \dodoi{10.1088/1674-4527/9/1/004}

\bibitem[{{Genel} {et~al.}(2015){Genel}, {Fall}, {Hernquist}, {Vogelsberger},
  {Snyder}, {Rodriguez-Gomez}, {Sijacki}, \& {Springel}}]{2015ApJ...804L..40G}
{Genel}, S., {Fall}, S.~M., {Hernquist}, L., {et~al.} 2015, \apjl, 804, L40,
  \dodoi{10.1088/2041-8205/804/2/L40}

\bibitem[{{Heavens} {et~al.}(2000){Heavens}, {Refregier}, \&
  {Heymans}}]{2000MNRAS.319..649H}
{Heavens}, A., {Refregier}, A., \& {Heymans}, C. 2000, \mnras, 319, 649,
  \dodoi{10.1046/j.1365-8711.2000.03907.x}

\bibitem[{{Hirata} {et~al.}(2007){Hirata}, {Mandelbaum}, {Ishak}, {Seljak},
  {Nichol}, {Pimbblet}, {Ross}, \& {Wake}}]{2007MNRAS.381.1197H}
{Hirata}, C.~M., {Mandelbaum}, R., {Ishak}, M., {et~al.} 2007, \mnras, 381,
  1197, \dodoi{10.1111/j.1365-2966.2007.12312.x}

\bibitem[{{Hirata} \& {Seljak}(2004)}]{2004PhRvD..70f3526H}
{Hirata}, C.~M., \& {Seljak}, U. 2004, \prd, 70, 063526,
  \dodoi{10.1103/PhysRevD.70.063526}

\bibitem[{{Hirata} {et~al.}(2004){Hirata}, {Mandelbaum}, {Seljak}, {Guzik},
  {Padmanabhan}, {Blake}, {Brinkmann}, {Bud{\'a}vari}, {Connolly}, {Csabai},
  {Scranton}, \& {Szalay}}]{2004MNRAS.353..529H}
{Hirata}, C.~M., {Mandelbaum}, R., {Seljak}, U., {et~al.} 2004, \mnras, 353,
  529, \dodoi{10.1111/j.1365-2966.2004.08090.x}

\bibitem[{{Hoffmann} {et~al.}(2022){Hoffmann}, {Secco}, {Blazek}, {Crocce},
  {Tallada-Cresp{\'\i}}, {Samuroff}, {Prat}, {Carretero}, {Fosalba},
  {Gazta{\~n}aga}, {Castander}, \& {DES Collaboration}}]{2022PhRvD.106l3510H}
{Hoffmann}, K., {Secco}, L.~F., {Blazek}, J., {et~al.} 2022, \prd, 106, 123510,
  \dodoi{10.1103/PhysRevD.106.123510}

\bibitem[{{Jagvaral} {et~al.}(2022){Jagvaral}, {Singh}, \&
  {Mandelbaum}}]{2022MNRAS.514.1021J}
{Jagvaral}, Y., {Singh}, S., \& {Mandelbaum}, R. 2022, \mnras, 514, 1021,
  \dodoi{10.1093/mnras/stac1424}

\bibitem[{{Jing}(2002)}]{2002MNRAS.335L..89J}
{Jing}, Y.~P. 2002, \mnras, 335, L89, \dodoi{10.1046/j.1365-8711.2002.05899.x}

\bibitem[{{Jing} {et~al.}(1995){Jing}, {Mo}, {Borner}, \&
  {Fang}}]{1995MNRAS.276..417J}
{Jing}, Y.~P., {Mo}, H.~J., {Borner}, G., \& {Fang}, L.~Z. 1995, \mnras, 276,
  417, \dodoi{10.1093/mnras/276.2.417}

\bibitem[{{Jing} \& {Suto}(2002)}]{2002ApJ...574..538J}
{Jing}, Y.~P., \& {Suto}, Y. 2002, \apj, 574, 538, \dodoi{10.1086/341065}

\bibitem[{{Katz}(1991)}]{1991ApJ...368..325K}
{Katz}, N. 1991, \apj, 368, 325, \dodoi{10.1086/169696}

\bibitem[{{Khandai} {et~al.}(2015){Khandai}, {Di Matteo}, {Croft}, {Wilkins},
  {Feng}, {Tucker}, {DeGraf}, \& {Liu}}]{2015MNRAS.450.1349K}
{Khandai}, N., {Di Matteo}, T., {Croft}, R., {et~al.} 2015, \mnras, 450, 1349,
  \dodoi{10.1093/mnras/stv627}

\bibitem[{{Kirk} {et~al.}(2012){Kirk}, {Rassat}, {Host}, \&
  {Bridle}}]{2012MNRAS.424.1647K}
{Kirk}, D., {Rassat}, A., {Host}, O., \& {Bridle}, S. 2012, \mnras, 424, 1647,
  \dodoi{10.1111/j.1365-2966.2012.21099.x}

\bibitem[{{Kirk} {et~al.}(2015){Kirk}, {Brown}, {Hoekstra}, {Joachimi},
  {Kitching}, {Mandelbaum}, {Sif{\'o}n}, {Cacciato}, {Choi}, {Kiessling},
  {Leonard}, {Rassat}, \& {Sch{\"a}fer}}]{2015SSRv..193..139K}
{Kirk}, D., {Brown}, M.~L., {Hoekstra}, H., {et~al.} 2015, \ssr, 193, 139,
  \dodoi{10.1007/s11214-015-0213-4}

\bibitem[{{Kogai} {et~al.}(2018){Kogai}, {Matsubara}, {Nishizawa}, \&
  {Urakawa}}]{2018JCAP...08..014K}
{Kogai}, K., {Matsubara}, T., {Nishizawa}, A.~J., \& {Urakawa}, Y. 2018, \jcap,
  2018, 014, \dodoi{10.1088/1475-7516/2018/08/014}

\bibitem[{{Krause} {et~al.}(2016){Krause}, {Eifler}, \&
  {Blazek}}]{2016MNRAS.456..207K}
{Krause}, E., {Eifler}, T., \& {Blazek}, J. 2016, \mnras, 456, 207,
  \dodoi{10.1093/mnras/stv2615}

\bibitem[{{Kurita} \& {Takada}(2023)}]{2023arXiv230202925K}
{Kurita}, T., \& {Takada}, M. 2023, arXiv e-prints, arXiv:2302.02925,
  \dodoi{10.48550/arXiv.2302.02925}

\bibitem[{{Landy} \& {Szalay}(1993)}]{1993ApJ...412...64L}
{Landy}, S.~D., \& {Szalay}, A.~S. 1993, \apj, 412, 64, \dodoi{10.1086/172900}

\bibitem[{{Li} {et~al.}(2013){Li}, {Jing}, {Faltenbacher}, \&
  {Wang}}]{2013ApJ...770L..12L}
{Li}, C., {Jing}, Y.~P., {Faltenbacher}, A., \& {Wang}, J. 2013, \apjl, 770,
  L12, \dodoi{10.1088/2041-8205/770/1/L12}

\bibitem[{{Mandelbaum} {et~al.}(2006){Mandelbaum}, {Hirata}, {Ishak}, {Seljak},
  \& {Brinkmann}}]{2006MNRAS.367..611M}
{Mandelbaum}, R., {Hirata}, C.~M., {Ishak}, M., {Seljak}, U., \& {Brinkmann},
  J. 2006, \mnras, 367, 611, \dodoi{10.1111/j.1365-2966.2005.09946.x}

\bibitem[{{Marinacci} {et~al.}(2018){Marinacci}, {Vogelsberger}, {Pakmor},
  {Torrey}, {Springel}, {Hernquist}, {Nelson}, {Weinberger}, {Pillepich},
  {Naiman}, \& {Genel}}]{2018MNRAS.480.5113M}
{Marinacci}, F., {Vogelsberger}, M., {Pakmor}, R., {et~al.} 2018, \mnras, 480,
  5113, \dodoi{10.1093/mnras/sty2206}

\bibitem[{{Naiman} {et~al.}(2018){Naiman}, {Pillepich}, {Springel},
  {Ramirez-Ruiz}, {Torrey}, {Vogelsberger}, {Pakmor}, {Nelson}, {Marinacci},
  {Hernquist}, {Weinberger}, \& {Genel}}]{2018MNRAS.477.1206N}
{Naiman}, J.~P., {Pillepich}, A., {Springel}, V., {et~al.} 2018, \mnras, 477,
  1206, \dodoi{10.1093/mnras/sty618}

\bibitem[{{Nelson} {et~al.}(2018){Nelson}, {Pillepich}, {Springel},
  {Weinberger}, {Hernquist}, {Pakmor}, {Genel}, {Torrey}, {Vogelsberger},
  {Kauffmann}, {Marinacci}, \& {Naiman}}]{2018MNRAS.475..624N}
{Nelson}, D., {Pillepich}, A., {Springel}, V., {et~al.} 2018, \mnras, 475, 624,
  \dodoi{10.1093/mnras/stx3040}

\bibitem[{{Nelson} {et~al.}(2019){Nelson}, {Springel}, {Pillepich},
  {Rodriguez-Gomez}, {Torrey}, {Genel}, {Vogelsberger}, {Pakmor}, {Marinacci},
  {Weinberger}, {Kelley}, {Lovell}, {Diemer}, \&
  {Hernquist}}]{2019ComAC...6....2N}
{Nelson}, D., {Springel}, V., {Pillepich}, A., {et~al.} 2019, Computational
  Astrophysics and Cosmology, 6, 2, \dodoi{10.1186/s40668-019-0028-x}

\bibitem[{{Okumura} \& {Jing}(2009)}]{2009ApJ...694L..83O}
{Okumura}, T., \& {Jing}, Y.~P. 2009, \apjl, 694, L83,
  \dodoi{10.1088/0004-637X/694/1/L83}

\bibitem[{{Okumura} {et~al.}(2009){Okumura}, {Jing}, \&
  {Li}}]{2009ApJ...694..214O}
{Okumura}, T., {Jing}, Y.~P., \& {Li}, C. 2009, \apj, 694, 214,
  \dodoi{10.1088/0004-637X/694/1/214}

\bibitem[{{Okumura} \& {Taruya}(2020)}]{2020MNRAS.493L.124O}
{Okumura}, T., \& {Taruya}, A. 2020, \mnras, 493, L124,
  \dodoi{10.1093/mnrasl/slaa024}

\bibitem[{{Okumura} \& {Taruya}(2022)}]{2022PhRvD.106d3523O}
---. 2022, \prd, 106, 043523, \dodoi{10.1103/PhysRevD.106.043523}

\bibitem[{{Okumura} \& {Taruya}(2023)}]{2023ApJ...945L..30O}
---. 2023, \apjl, 945, L30, \dodoi{10.3847/2041-8213/acbf48}

\bibitem[{{Okumura} {et~al.}(2020){Okumura}, {Taruya}, \&
  {Nishimichi}}]{2020MNRAS.494..694O}
{Okumura}, T., {Taruya}, A., \& {Nishimichi}, T. 2020, \mnras, 494, 694,
  \dodoi{10.1093/mnras/staa718}

\bibitem[{{Pen} {et~al.}(2000){Pen}, {Lee}, \& {Seljak}}]{2000ApJ...543L.107P}
{Pen}, U.-L., {Lee}, J., \& {Seljak}, U. 2000, \apjl, 543, L107,
  \dodoi{10.1086/317273}

\bibitem[{{Pillepich} {et~al.}(2018{\natexlab{a}}){Pillepich}, {Nelson},
  {Hernquist}, {Springel}, {Pakmor}, {Torrey}, {Weinberger}, {Genel}, {Naiman},
  {Marinacci}, \& {Vogelsberger}}]{2018MNRAS.475..648P}
{Pillepich}, A., {Nelson}, D., {Hernquist}, L., {et~al.} 2018{\natexlab{a}},
  \mnras, 475, 648, \dodoi{10.1093/mnras/stx3112}

\bibitem[{{Pillepich} {et~al.}(2018{\natexlab{b}}){Pillepich}, {Springel},
  {Nelson}, {Genel}, {Naiman}, {Pakmor}, {Hernquist}, {Torrey}, {Vogelsberger},
  {Weinberger}, \& {Marinacci}}]{2018MNRAS.473.4077P}
{Pillepich}, A., {Springel}, V., {Nelson}, D., {et~al.} 2018{\natexlab{b}},
  \mnras, 473, 4077, \dodoi{10.1093/mnras/stx2656}

\bibitem[{{Planck Collaboration} {et~al.}(2016){Planck Collaboration}, {Ade},
  {Aghanim}, {Arnaud}, {Ashdown}, {Aumont}, {Baccigalupi}, {Banday},
  {Barreiro}, {Bartlett}, {Bartolo}, {Battaner}, {Battye}, {Benabed},
  {Beno{\^\i}t}, {Benoit-L{\'e}vy}, {Bernard}, {Bersanelli}, {Bielewicz},
  {Bock}, {Bonaldi}, {Bonavera}, {Bond}, {Borrill}, {Bouchet}, {Boulanger},
  {Bucher}, {Burigana}, {Butler}, {Calabrese}, {Cardoso}, {Catalano},
  {Challinor}, {Chamballu}, {Chary}, {Chiang}, {Chluba}, {Christensen},
  {Church}, {Clements}, {Colombi}, {Colombo}, {Combet}, {Coulais}, {Crill},
  {Curto}, {Cuttaia}, {Danese}, {Davies}, {Davis}, {de Bernardis}, {de Rosa},
  {de Zotti}, {Delabrouille}, {D{\'e}sert}, {Di Valentino}, {Dickinson},
  {Diego}, {Dolag}, {Dole}, {Donzelli}, {Dor{\'e}}, {Douspis}, {Ducout},
  {Dunkley}, {Dupac}, {Efstathiou}, {Elsner}, {En{\ss}lin}, {Eriksen},
  {Farhang}, {Fergusson}, {Finelli}, {Forni}, {Frailis}, {Fraisse},
  {Franceschi}, {Frejsel}, {Galeotta}, {Galli}, {Ganga}, {Gauthier}, {Gerbino},
  {Ghosh}, {Giard}, {Giraud-H{\'e}raud}, {Giusarma}, {Gjerl{\o}w},
  {Gonz{\'a}lez-Nuevo}, {G{\'o}rski}, {Gratton}, {Gregorio}, {Gruppuso},
  {Gudmundsson}, {Hamann}, {Hansen}, {Hanson}, {Harrison}, {Helou},
  {Henrot-Versill{\'e}}, {Hern{\'a}ndez-Monteagudo}, {Herranz}, {Hildebrandt},
  {Hivon}, {Hobson}, {Holmes}, {Hornstrup}, {Hovest}, {Huang}, {Huffenberger},
  {Hurier}, {Jaffe}, {Jaffe}, {Jones}, {Juvela}, {Keih{\"a}nen}, {Keskitalo},
  {Kisner}, {Kneissl}, {Knoche}, {Knox}, {Kunz}, {Kurki-Suonio}, {Lagache},
  {L{\"a}hteenm{\"a}ki}, {Lamarre}, {Lasenby}, {Lattanzi}, {Lawrence}, {Leahy},
  {Leonardi}, {Lesgourgues}, {Levrier}, {Lewis}, {Liguori}, {Lilje},
  {Linden-V{\o}rnle}, {L{\'o}pez-Caniego}, {Lubin}, {Mac{\'\i}as-P{\'e}rez},
  {Maggio}, {Maino}, {Mandolesi}, {Mangilli}, {Marchini}, {Maris}, {Martin},
  {Martinelli}, {Mart{\'\i}nez-Gonz{\'a}lez}, {Masi}, {Matarrese}, {McGehee},
  {Meinhold}, {Melchiorri}, {Melin}, {Mendes}, {Mennella}, {Migliaccio},
  {Millea}, {Mitra}, {Miville-Desch{\^e}nes}, {Moneti}, {Montier}, {Morgante},
  {Mortlock}, {Moss}, {Munshi}, {Murphy}, {Naselsky}, {Nati}, {Natoli},
  {Netterfield}, {N{\o}rgaard-Nielsen}, {Noviello}, {Novikov}, {Novikov},
  {Oxborrow}, {Paci}, {Pagano}, {Pajot}, {Paladini}, {Paoletti}, {Partridge},
  {Pasian}, {Patanchon}, {Pearson}, {Perdereau}, {Perotto}, {Perrotta},
  {Pettorino}, {Piacentini}, {Piat}, {Pierpaoli}, {Pietrobon}, {Plaszczynski},
  {Pointecouteau}, {Polenta}, {Popa}, {Pratt}, {Pr{\'e}zeau}, {Prunet},
  {Puget}, {Rachen}, {Reach}, {Rebolo}, {Reinecke}, {Remazeilles}, {Renault},
  {Renzi}, {Ristorcelli}, {Rocha}, {Rosset}, {Rossetti}, {Roudier},
  {Rouill{\'e} d'Orfeuil}, {Rowan-Robinson}, {Rubi{\~n}o-Mart{\'\i}n},
  {Rusholme}, {Said}, {Salvatelli}, {Salvati}, {Sandri}, {Santos},
  {Savelainen}, {Savini}, {Scott}, {Seiffert}, {Serra}, {Shellard}, {Spencer},
  {Spinelli}, {Stolyarov}, {Stompor}, {Sudiwala}, {Sunyaev}, {Sutton},
  {Suur-Uski}, {Sygnet}, {Tauber}, {Terenzi}, {Toffolatti}, {Tomasi},
  {Tristram}, {Trombetti}, {Tucci}, {Tuovinen}, {T{\"u}rler}, {Umana},
  {Valenziano}, {Valiviita}, {Van Tent}, {Vielva}, {Villa}, {Wade}, {Wandelt},
  {Wehus}, {White}, {White}, {Wilkinson}, {Yvon}, {Zacchei}, \&
  {Zonca}}]{2016A&A...594A..13P}
{Planck Collaboration}, {Ade}, P.~A.~R., {Aghanim}, N., {et~al.} 2016, \aap,
  594, A13, \dodoi{10.1051/0004-6361/201525830}

\bibitem[{{Reid} {et~al.}(2016){Reid}, {Ho}, {Padmanabhan}, {Percival},
  {Tinker}, {Tojeiro}, {White}, {Eisenstein}, {Maraston}, {Ross},
  {S{\'a}nchez}, {Schlegel}, {Sheldon}, {Strauss}, {Thomas}, {Wake}, {Beutler},
  {Bizyaev}, {Bolton}, {Brownstein}, {Chuang}, {Dawson}, {Harding}, {Kitaura},
  {Leauthaud}, {Masters}, {McBride}, {More}, {Olmstead}, {Oravetz}, {Nuza},
  {Pan}, {Parejko}, {Pforr}, {Prada}, {Rodr{\'\i}guez-Torres},
  {Salazar-Albornoz}, {Samushia}, {Schneider}, {Sc{\'o}ccola}, {Simmons}, \&
  {Vargas-Magana}}]{2016MNRAS.455.1553R}
{Reid}, B., {Ho}, S., {Padmanabhan}, N., {et~al.} 2016, \mnras, 455, 1553,
  \dodoi{10.1093/mnras/stv2382}

\bibitem[{{Rodriguez} {et~al.}(2022){Rodriguez}, {Merch{\'a}n}, \&
  {Artale}}]{2022MNRAS.514.1077R}
{Rodriguez}, F., {Merch{\'a}n}, M., \& {Artale}, M.~C. 2022, \mnras, 514, 1077,
  \dodoi{10.1093/mnras/stac1428}

\bibitem[{{Rodriguez-Gomez} {et~al.}(2015){Rodriguez-Gomez}, {Genel},
  {Vogelsberger}, {Sijacki}, {Pillepich}, {Sales}, {Torrey}, {Snyder},
  {Nelson}, {Springel}, {Ma}, \& {Hernquist}}]{2015MNRAS.449...49R}
{Rodriguez-Gomez}, V., {Genel}, S., {Vogelsberger}, M., {et~al.} 2015, \mnras,
  449, 49, \dodoi{10.1093/mnras/stv264}

\bibitem[{{Rodriguez-Gomez} {et~al.}(2016){Rodriguez-Gomez}, {Pillepich},
  {Sales}, {Genel}, {Vogelsberger}, {Zhu}, {Wellons}, {Nelson}, {Torrey},
  {Springel}, {Ma}, \& {Hernquist}}]{2016MNRAS.458.2371R}
{Rodriguez-Gomez}, V., {Pillepich}, A., {Sales}, L.~V., {et~al.} 2016, \mnras,
  458, 2371, \dodoi{10.1093/mnras/stw456}

\bibitem[{{Samuroff} {et~al.}(2022){Samuroff}, {Mandelbaum}, {Blazek},
  {Campos}, {MacCrann}, {Zacharegkas}, {Amon}, {Prat}, {Singh}, {Elvin-Poole},
  {Ross}, {Alarcon}, {Baxter}, {Bechtol}, {Becker}, {Bernstein}, {Carnero
  Rosell}, {Carrasco Kind}, {Cawthon}, {Chang}, {Chen}, {Choi}, {Crocce},
  {Davis}, {DeRose}, {Dodelson}, {Doux}, {Drlica-Wagner}, {Eckert}, {Everett},
  {Fert{\'e}}, {Gatti}, {Giannini}, {Gruen}, {Gruendl}, {Harrison}, {Herner},
  {Huff}, {Jarvis}, {Kuropatkin}, {Leget}, {Lemos}, {McCullough}, {Myles},
  {Navarro-Alsina}, {Pandey}, {Porredon}, {Raveri}, {Rodriguez-Monroy},
  {Rollins}, {Roodman}, {Rossi}, {Rykoff}, {S{\'a}nchez}, {Secco},
  {Sevilla-Noarbe}, {Sheldon}, {Shin}, {Troxel}, {Tutusaus}, {Weaverdyck},
  {Yanny}, {Yin}, {Zhang}, {Aguena}, {Alves}, {Annis}, {Bacon}, {Bertin},
  {Bocquet}, {Brooks}, {Burke}, {Carretero}, {Costanzi}, {da Costa}, {Pereira},
  {De Vicente}, {Desai}, {Diehl}, {Dietrich}, {Doel}, {Ferrero}, {Flaugher},
  {Frieman}, {Garc{\'\i}a-Bellido}, {Hinton}, {Hollowood}, {Honscheid},
  {James}, {Kuehn}, {Lahav}, {Marshall}, {Melchior}, {Mena-Fern{\'a}ndez},
  {Menanteau}, {Miquel}, {Newman}, {Palmese}, {Pieres}, {Plazas Malag{\'o}n},
  {Sanchez}, {Scarpine}, {Smith}, {Suchyta}, {Swanson}, {Tarle}, \&
  {To}}]{2022arXiv221211319S}
{Samuroff}, S., {Mandelbaum}, R., {Blazek}, J., {et~al.} 2022, arXiv e-prints,
  arXiv:2212.11319, \dodoi{10.48550/arXiv.2212.11319}

\bibitem[{{Schaye} {et~al.}(2010){Schaye}, {Dalla Vecchia}, {Booth}, {Wiersma},
  {Theuns}, {Haas}, {Bertone}, {Duffy}, {McCarthy}, \& {van de
  Voort}}]{2010MNRAS.402.1536S}
{Schaye}, J., {Dalla Vecchia}, C., {Booth}, C.~M., {et~al.} 2010, \mnras, 402,
  1536, \dodoi{10.1111/j.1365-2966.2009.16029.x}

\bibitem[{{Schaye} {et~al.}(2015){Schaye}, {Crain}, {Bower}, {Furlong},
  {Schaller}, {Theuns}, {Dalla Vecchia}, {Frenk}, {McCarthy}, {Helly},
  {Jenkins}, {Rosas-Guevara}, {White}, {Baes}, {Booth}, {Camps}, {Navarro},
  {Qu}, {Rahmati}, {Sawala}, {Thomas}, \& {Trayford}}]{2015MNRAS.446..521S}
{Schaye}, J., {Crain}, R.~A., {Bower}, R.~G., {et~al.} 2015, \mnras, 446, 521,
  \dodoi{10.1093/mnras/stu2058}

\bibitem[{{Schmidt} {et~al.}(2015){Schmidt}, {Chisari}, \&
  {Dvorkin}}]{2015JCAP...10..032S}
{Schmidt}, F., {Chisari}, N.~E., \& {Dvorkin}, C. 2015, \jcap, 2015, 032,
  \dodoi{10.1088/1475-7516/2015/10/032}

\bibitem[{{Schneider} \& {Bridle}(2010)}]{2010MNRAS.402.2127S}
{Schneider}, M.~D., \& {Bridle}, S. 2010, \mnras, 402, 2127,
  \dodoi{10.1111/j.1365-2966.2009.15956.x}

\bibitem[{{Sheng} {et~al.}(2023){Sheng}, {Yu}, {Li}, {Liao}, {Du}, {Wang},
  {Wang}, {Xu}, {Genel}, \& {Irodotou}}]{2023ApJ...943..128S}
{Sheng}, M.-J., {Yu}, H.-R., {Li}, S., {et~al.} 2023, \apj, 943, 128,
  \dodoi{10.3847/1538-4357/acae92}

\bibitem[{{Singh} {et~al.}(2015){Singh}, {Mandelbaum}, \&
  {More}}]{2015MNRAS.450.2195S}
{Singh}, S., {Mandelbaum}, R., \& {More}, S. 2015, \mnras, 450, 2195,
  \dodoi{10.1093/mnras/stv778}

\bibitem[{{Springel}(2010)}]{2010MNRAS.401..791S}
{Springel}, V. 2010, \mnras, 401, 791, \dodoi{10.1111/j.1365-2966.2009.15715.x}

\bibitem[{{Springel} {et~al.}(2001){Springel}, {White}, {Tormen}, \&
  {Kauffmann}}]{2001MNRAS.328..726S}
{Springel}, V., {White}, S. D.~M., {Tormen}, G., \& {Kauffmann}, G. 2001,
  \mnras, 328, 726, \dodoi{10.1046/j.1365-8711.2001.04912.x}

\bibitem[{{Springel} {et~al.}(2018){Springel}, {Pakmor}, {Pillepich},
  {Weinberger}, {Nelson}, {Hernquist}, {Vogelsberger}, {Genel}, {Torrey},
  {Marinacci}, \& {Naiman}}]{2018MNRAS.475..676S}
{Springel}, V., {Pakmor}, R., {Pillepich}, A., {et~al.} 2018, \mnras, 475, 676,
  \dodoi{10.1093/mnras/stx3304}

\bibitem[{{Tenneti} {et~al.}(2016){Tenneti}, {Mandelbaum}, \& {Di
  Matteo}}]{2016MNRAS.462.2668T}
{Tenneti}, A., {Mandelbaum}, R., \& {Di Matteo}, T. 2016, \mnras, 462, 2668,
  \dodoi{10.1093/mnras/stw1823}

\bibitem[{{Tenneti} {et~al.}(2014){Tenneti}, {Mandelbaum}, {Di Matteo}, {Feng},
  \& {Khandai}}]{2014MNRAS.441..470T}
{Tenneti}, A., {Mandelbaum}, R., {Di Matteo}, T., {Feng}, Y., \& {Khandai}, N.
  2014, \mnras, 441, 470, \dodoi{10.1093/mnras/stu586}

\bibitem[{{The Dark Energy Survey Collaboration}(2005)}]{2005astro.ph.10346T}
{The Dark Energy Survey Collaboration}. 2005, arXiv e-prints, astro,
  \dodoi{10.48550/arXiv.astro-ph/0510346}

\bibitem[{{Velliscig} {et~al.}(2015){Velliscig}, {Cacciato}, {Schaye}, {Crain},
  {Bower}, {van Daalen}, {Dalla Vecchia}, {Frenk}, {Furlong}, {McCarthy},
  {Schaller}, \& {Theuns}}]{2015MNRAS.453..721V}
{Velliscig}, M., {Cacciato}, M., {Schaye}, J., {et~al.} 2015, \mnras, 453, 721,
  \dodoi{10.1093/mnras/stv1690}

\bibitem[{{Weinberger} {et~al.}(2017){Weinberger}, {Springel}, {Hernquist},
  {Pillepich}, {Marinacci}, {Pakmor}, {Nelson}, {Genel}, {Vogelsberger},
  {Naiman}, \& {Torrey}}]{2017MNRAS.465.3291W}
{Weinberger}, R., {Springel}, V., {Hernquist}, L., {et~al.} 2017, \mnras, 465,
  3291, \dodoi{10.1093/mnras/stw2944}

\bibitem[{{Xia} {et~al.}(2017){Xia}, {Kang}, {Wang}, {Luo}, {Yang}, {Jing},
  {Wang}, \& {Mo}}]{2017ApJ...848...22X}
{Xia}, Q., {Kang}, X., {Wang}, P., {et~al.} 2017, \apj, 848, 22,
  \dodoi{10.3847/1538-4357/aa8d17}

\bibitem[{{Xu} {et~al.}(2023{\natexlab{a}}){Xu}, {Jing}, \&
  {Gao}}]{2023arXiv230204230X}
{Xu}, K., {Jing}, Y.~P., \& {Gao}, H. 2023{\natexlab{a}}, arXiv e-prints,
  arXiv:2302.04230, \dodoi{10.48550/arXiv.2302.04230}

\bibitem[{{Xu} {et~al.}(2023{\natexlab{b}}){Xu}, {Jing}, {Zhao}, \&
  {Cuesta}}]{2023arXiv230609407X}
{Xu}, K., {Jing}, Y.~P., {Zhao}, G.-B., \& {Cuesta}, A.~J. 2023{\natexlab{b}},
  arXiv e-prints, arXiv:2306.09407, \dodoi{10.48550/arXiv.2306.09407}

\bibitem[{{Xu} {et~al.}(2023{\natexlab{c}}){Xu}, {Jing}, {Zheng}, \&
  {Gao}}]{2023ApJ...944..200X}
{Xu}, K., {Jing}, Y.~P., {Zheng}, Y., \& {Gao}, H. 2023{\natexlab{c}}, \apj,
  944, 200, \dodoi{10.3847/1538-4357/acb13e}

\bibitem[{{Yang} {et~al.}(2006){Yang}, {van den Bosch}, {Mo}, {Mao}, {Kang},
  {Weinmann}, {Guo}, \& {Jing}}]{2006MNRAS.369.1293Y}
{Yang}, X., {van den Bosch}, F.~C., {Mo}, H.~J., {et~al.} 2006, \mnras, 369,
  1293, \dodoi{10.1111/j.1365-2966.2006.10373.x}

\bibitem[{{Yao} {et~al.}(2020){Yao}, {Shan}, {Zhang}, {Kneib}, \&
  {Jullo}}]{2020ApJ...904..135Y}
{Yao}, J., {Shan}, H., {Zhang}, P., {Kneib}, J.-P., \& {Jullo}, E. 2020, \apj,
  904, 135, \dodoi{10.3847/1538-4357/abc175}

\bibitem[{{York} {et~al.}(2000){York}, {Adelman}, {Anderson}, {Anderson},
  {Annis}, {Bahcall}, {Bakken}, {Barkhouser}, {Bastian}, {Berman}, {Boroski},
  {Bracker}, {Briegel}, {Briggs}, {Brinkmann}, {Brunner}, {Burles}, {Carey},
  {Carr}, {Castander}, {Chen}, {Colestock}, {Connolly}, {Crocker}, {Csabai},
  {Czarapata}, {Davis}, {Doi}, {Dombeck}, {Eisenstein}, {Ellman}, {Elms},
  {Evans}, {Fan}, {Federwitz}, {Fiscelli}, {Friedman}, {Frieman}, {Fukugita},
  {Gillespie}, {Gunn}, {Gurbani}, {de Haas}, {Haldeman}, {Harris}, {Hayes},
  {Heckman}, {Hennessy}, {Hindsley}, {Holm}, {Holmgren}, {Huang}, {Hull},
  {Husby}, {Ichikawa}, {Ichikawa}, {Ivezi{\'c}}, {Kent}, {Kim}, {Kinney},
  {Klaene}, {Kleinman}, {Kleinman}, {Knapp}, {Korienek}, {Kron}, {Kunszt},
  {Lamb}, {Lee}, {Leger}, {Limmongkol}, {Lindenmeyer}, {Long}, {Loomis},
  {Loveday}, {Lucinio}, {Lupton}, {MacKinnon}, {Mannery}, {Mantsch}, {Margon},
  {McGehee}, {McKay}, {Meiksin}, {Merelli}, {Monet}, {Munn}, {Narayanan},
  {Nash}, {Neilsen}, {Neswold}, {Newberg}, {Nichol}, {Nicinski}, {Nonino},
  {Okada}, {Okamura}, {Ostriker}, {Owen}, {Pauls}, {Peoples}, {Peterson},
  {Petravick}, {Pier}, {Pope}, {Pordes}, {Prosapio}, {Rechenmacher}, {Quinn},
  {Richards}, {Richmond}, {Rivetta}, {Rockosi}, {Ruthmansdorfer}, {Sandford},
  {Schlegel}, {Schneider}, {Sekiguchi}, {Sergey}, {Shimasaku}, {Siegmund},
  {Smee}, {Smith}, {Snedden}, {Stone}, {Stoughton}, {Strauss}, {Stubbs},
  {SubbaRao}, {Szalay}, {Szapudi}, {Szokoly}, {Thakar}, {Tremonti}, {Tucker},
  {Uomoto}, {Vanden Berk}, {Vogeley}, {Waddell}, {Wang}, {Watanabe},
  {Weinberg}, {Yanny}, {Yasuda}, \& {SDSS Collaboration}}]{2000AJ....120.1579Y}
{York}, D.~G., {Adelman}, J., {Anderson}, John~E., J., {et~al.} 2000, \aj, 120,
  1579, \dodoi{10.1086/301513}

\end{thebibliography}
\bibliographystyle{aasjournal}

%% This command is needed to show the entire author+affiliation list when
%% the collaboration and author truncation commands are used.  It has to
%% go at the end of the manuscript.
%\allauthors

%% Include this line if you are using the \added, \replaced, \deleted
%% commands to see a summary list of all changes at the end of the article.
%\listofchanges

\appendix
\restartappendixnumbering
\section{Generating random numbers for TSE distribution}\label{sec:A}
To generate random numbers with TSE distribution from a uniform distribution $U(0,1)$, we can use the following method. Let $X$ be a random variable following $U(0,1)$, and we want to find a function $y$ such that $Y=y(X)$ follows a TSE distribution. The cumulative distribution functions of $X$ and $Y$ have the relation $F_X(x)=F_Y(y)$, which leads to the following equation:
\begin{align}
    x&=\int_{0}^{y}(Ae^{-y'/\tau}+B)dy'\notag\\
     &=\tau A(1-e^{-y/\tau})+By\,\,.
\end{align}
Then, We can solve for $y(x)$ by inverting the above equation as follows:
\begin{equation}
    z = \frac{A}{B}\exp\left(\frac{A\tau-x}{B\tau}\right)\,\,,
\end{equation}
\begin{equation}
    y=\tau W_0(z)-\frac{A\tau-x}{B}\,\,.
\end{equation}
Finally, we can generate random numbers following the TSE distribution by computing $y(X)$.

\section{The sum of exponential functions}\label{sec:B}
\begin{figure*}[h]
    \plotone{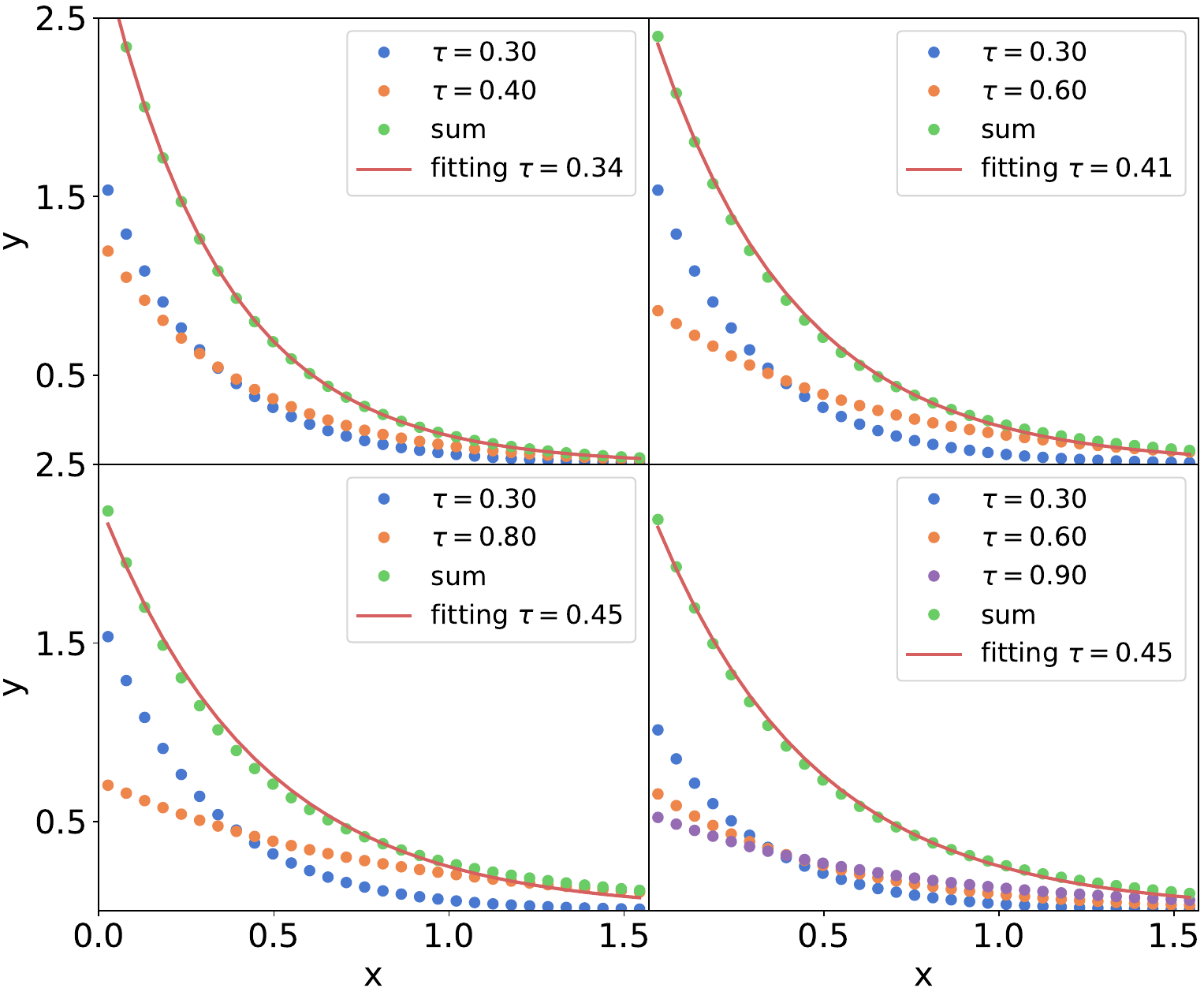}
    \caption{We demonstrate that the sum of exponential functions with similar exponents $\tau$ can also be effectively approximated by a single exponential function. All exponential functions are normalized within the interval $[0,\pi/2]$. In cases where $n$ exponential functions are summed up, all the exponential functions being summed are divided by $n$.}
    \label{fig:test_exp}
\end{figure*}
We test the self-consistency of our TSE model by demonstrating that the sum of exponential functions with close exponents $\tau$ can also be well fitted by the exponential function. Figure \ref{fig:test_exp} shows the cases where two or three exponential functions with different exponents $\tau$ are summed up, and we find that the sum can also be well fitted by the exponential function.

\end{document}